\documentclass[sigconf]{acmart}
\usepackage{cancel}
\usepackage{color}
\usepackage{svg}
\usepackage{xcolor}
\usepackage{graphicx}
\usepackage{enumitem}

\usepackage[section]{algorithm}
\usepackage{algorithmicx} \usepackage{algpseudocode}

\def\figcapup{\vspace{-1mm}}
\def\figcapdown{\vspace{-0mm}}

\newtheorem{fact}{Fact}
\newtheorem{assumption}{Assumption}

\def\-{\mbox{-}}

\def\lf{\lfloor}

\def\rf{\rfloor}

\def\Pr{\mathbf{Pr}}

\def\*{\star}

\newcommand{\red}[1]{}
\newcommand{\blue}[1]{}
\newcommand{\newred}[1]{}
\newcommand{\newblue}[1]{}

\renewcommand{\red}[1]{\textcolor{black}{#1}}
\renewcommand{\blue}[1]{\textcolor{blue}{#1}}
\renewcommand{\newred}[1]{\color{red}{#1}}
\renewcommand{\newblue}[1]{\color{black}{#1}}

\usepackage{xcolor}

\newcommand{\tony}[1]{}
\newcommand{\junhao}[1]{}
\newcommand{\hao}[1]{}
\newcommand{\todo}[1]{}

\newcommand{\DT}{\ensuremath{\mathit{CMY}}}
\newcommand{\CMY}{\ensuremath{\mathit{CMY}}}

\newcommand{\cmy}{\ensuremath{\mathit{CMY}}}

\newcommand{\DstT}{\ensuremath{\mathit{}{StcSlk\text{-}KwnDst}}}
\newcommand{\DstDT}{\ensuremath{\mathit{DynSlk\text{-}KwnDst}}}
\newcommand{\LnDstT}{\ensuremath{\mathit{StcSlk\text{-}LrnDst}}}
\newcommand{\LnDstDT}{\ensuremath{\mathit{DynSlk\text{-}LrnDst}}}
\newcommand{\US}{\ensuremath{\mathit{UniSlk}}}

\newcommand{\myqed}{\begin{flushright}$\qed$\end{flushright}
}

\copyrightyear{2020} 
\acmYear{2020} 
\setcopyright{acmlicensed}\acmConference[KDD '20]{Proceedings of the 26th ACM SIGKDD Conference on Knowledge Discovery and Data Mining}{August 23--27, 2020}{Virtual Event, CA, USA}
\acmBooktitle{Proceedings of the 26th ACM SIGKDD Conference on Knowledge Discovery and Data Mining (KDD '20), August 23--27, 2020, Virtual Event, CA, USA}
\acmPrice{15.00}
\acmDOI{10.1145/3394486.3403255}
\acmISBN{978-1-4503-7998-4/20/08}

\begin{document}
\fancyhead{}

\title{Learning Based Distributed Tracking}

\author{Hao WU}
\email{whw4@student.unimelb.edu.au}
\affiliation{
  \institution{The University of Melbourne}
  \city{Melbourne}
  \country{Australia}
}

\author{Junhao Gan}
\email{junhao.gan@unimelb.edu.au}
\affiliation{
 \institution{The University of Melbourne}
 \city{Melbourne}
 \country{Australia}
}

\author{Rui Zhang}
\email{rui.zhang@unimelb.edu.au}
\affiliation{
 \institution{The University of Melbourne}
 \city{Melbourne}
 \country{Australia}
}


\begin{abstract}
 Inspired by the great success of machine learning in the past decade, people have been thinking about
 the possibility of improving the theoretical results by exploring data distribution.
 In this paper, we revisit a fundamental problem called \emph{Distributed Tracking} (DT) 
 under an assumption that the data follows a certain (known or unknown) distribution, and propose a number \emph{data-dependent} algorithms with improved theoretical bounds. 
 Informally, in the DT problem, there is a coordinator and $k$ players, where the coordinator holds a threshold $N$ and each player has a counter.
 At each time stamp, at most one counter can be increased by one. The job of the coordinator is to capture the exact moment when the sum of all these $k$ counters reaches $N$. The goal is to minimise the communication cost. 
 While our first type of algorithms assume the concrete data distribution is \emph{known in advance}, our second type of algorithms can learn the distribution on the fly.  
 Both of the algorithms achieve a communication cost bounded by $O(k\log \log N)$ with high probability, improving the state-of-the-art \emph{data-independent} bound $O(k \log \frac{N}{k})$.
 We further propose a number of implementation optimisation heuristics to improve both efficiency and robustness of the algorithms.
 Finally, we conduct extensive experiments on three real datasets and four synthetic datasets. The experimental results show that the communication cost of our algorithms is as least as $20\%$ of that of the state-of-the-art algorithms. 
\end{abstract}




\maketitle

\begin{sloppy}

\section{Introduction}
{\newblue
	The great success of machine learning in the past decade has proven the assumption that data in practice follows certain patterns (e.g., either known or unknown distributions). 
	Such an assumption in turn becomes the base of those (machine learning) techniques.
Inspired by this, 
people have been thinking about the possibility to improve theoretical results on traditional problems with machine learning techniques. Successful progresses have been made on a wide range of problems, such as frequency estimation \cite{HIKV19, AIV19}, approximate membership \cite{M18}, combinatorial optimization \cite{KDZDS17, VFJ15, BPL0B17}, index structures \cite{KBCDP18}, and etc.   
The rationale behind these progresses is to explore and exploit the underlying distribution of the input data and to design \emph{data-dependent} algorithms customized for the data distribution.
In this paper, we design data-dependent algorithms, improving the state-of-the-art theoretical bounds, for solving the \emph{Distribution Tracking} (DT) problem~\cite{CMY11}. 

\vspace{1mm}
\noindent
{\bf The DT problem setting.}
In the DT problem, there is a coordinator and $k$ players (a.k.a. sites); between each player and the coordinator, there is a two-way communication channel. 
The coordinator holds a \emph{threshold} $N$, while the $i$-th player has a \emph{counter} $n_i$ initialized as $0$ for all $i \in \{1, 2, \ldots, k\}$. At each time stamp, there is \emph{at most one} (that means it can be none) player having its counter increased by one, conceptually representing an {item} arrives at the player.
The job of the coordinator is to raise an alarm at the \emph{exact moment} that the $N$-th item arrives (at some player), equivalently, the moment that the sum of the counters of all the players reaches $N$, i.e., $\sum_{i = 1}^k n_i = N$. 
The efficiency of an algorithm for solving the DT problem is measured by the communication cost, i.e., the total number of \emph{messages} that received and sent by the coordinator, where each message can only carry at most $O(1)$ \emph{words}.
}

{\newblue
	To solve the DT problem, a straightforward algorithm is to instruct each player to send a message to notify the coordinator for every increment on its counter. The communication cost of this algorithm is clearly $N$ (messages). However, such a communication cost is considered expensive, as $N$ is large in practice.
Existing work~\cite{CMY11} 
showed that the DT problem actually admits an algorithm (the $\cmy$ algorithm named after its authors) with a communication cost of $O(k\log \frac{N}{k})$. When $N$ is far larger than $k$, this algorithm consumes significantly less communication cost than the straightforward algorithm. 

\vspace{1mm}
\noindent
{\bf The $\cmy$ algorithm.} For the ease of explanation,  we introduce a simplified version of the state-of-the-art  $\cmy$ algorithm achieving the same communication bound. The simplified algorithm runs in \emph{rounds}. In each round, if $N < 4k$, run the straightforward algorithm with $O(N) = O(k)$ messages. Otherwise (i.e., $N \geq 4k$), the coordinator sends a \emph{slack} $s = \lf \frac{N}{2k} \rf$ to each player.
Each player sends a message to notify the coordinator, whenever its counter is increased by $s$ since its last communication with the coordinator. 
If the coordinator receives the $k$-th message, then it collects all the $k$ counter values $n_i$ from the players and calculate $N' = N - \sum_{i = 1}^k n_i$. If $N' = 0$, the coordinator raises the alarm. Otherwise, start a new round to solve a \emph{new} DT problem instance with $N = N'$ from scratch.
As it is easy to verified that in each round, the coordinator sends and receives $O(k)$ messages; and after each round, $N$ is decreased by a constant factor. Hence, there can be at most $O(\log \frac{N}{k})$ (with respect to the original $N$) rounds; the total communication cost is bounded by $O(k \log \frac{N}{k})$ messages.

\begin{figure}[t]
    \centering
    \hspace{-4mm}
    \includegraphics[width=.95\linewidth]{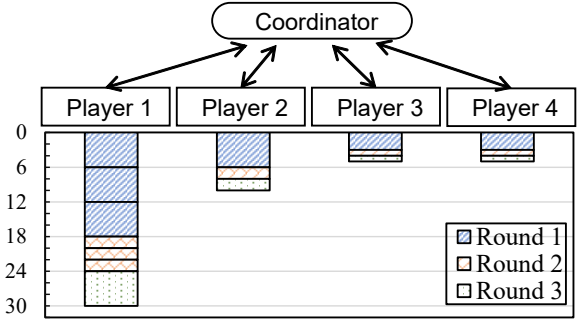}
    \vspace{-4mm}
    \caption{A running example of the simplified $\cmy$ algorithm with $N = 50$ and $k = 4$, where the counter increments in different rounds are highlighted with different colors and textures. The slack values in the first two rounds are $s=6$ and $s=2$, respectively; and in the third round, the algorithm switches to the straightforward algorithm.
    }
    \vspace{-5mm}
    \label{fig:example}
\end{figure}

\vspace{1mm}
\noindent
{\bf A running example.} Figure~\ref{fig:example} shows a running example of the DT algorithm on an instance with $N = 50$ and $k = 4$, 
In the first round, since $N > 4k = 16$, the coordinator sends a slack $s = \lf \frac{N}{2k} \rf = \lf \frac{50}{2*4} \rf = 6$ to each player. At the end of this round, Player 1 has received 18 counter increments and thus in total 3 messages have been sent from Player 1 to the coordinator, which were sent for every $s = 6$ increments. Likewise, Player 2 has sent a message as its counter is increased by $s =6$, while both of the counters of Players 3 and 4 are just increased by $3$: no messages were sent from them. The coordinator collects all the counters as soon as it receives the $k$-th (i.e., the fourth) messages and calculate $N' = 50 - (6*3 + 6 + 3 + 3) = 20$.   
Next, the coordinator starts a new round with a new instance with $N = 20$ and $k = 4$ from scratch, where $s  = \lf \frac{20}{2*4} \rf = 2$. As shown in the figure, at the end of this round, the counters of the four players have been increased by $6$, $2$, $1$ and $1$, respectively; and $N' = 10$. A new round with $N = 10$ and $k = 4$ is thus started, in which the algorithm switches to the straightforward algorithm (as $N < 4k$): a message is sent for each counter increment.
As for the communication cost, in each of the first two rounds, the coordinator sends $4$ messages for sending the slack $s$ to the four players, receives $4$ messages, sends $4$ messages for requesting the counter values, and finally receives $4$ messages for the counter values. Therefore, the communication cost in each of these two rounds is $16$. Plus the $4 + 10$ messages in the last round, the total communication cost is thus $16*2 + 14 = 46$.

\vspace{1mm}
\noindent
{\bf Exploiting the counter increment distribution.} 
In order to provide a worst-case communication bound, the $\cmy$ algorithm has to
be pessimistic and conservative: it makes  
no assumption on the data distribution; and 
only identical slacks $s$ can be sent to all the players in a round. 
Such pessimism and conservation, 
unfortunately, prevents the $\cmy$ algorithm from reducing the communication by exploiting the data distribution. In particular, the data distribution we mean here is the \emph{Multinomial Distribution of the counter increments}, more specifically, the \emph{probability distribution} of a counter increment happening in the players.
For the example shown in Figure~\ref{fig:example}, the probability distribution of the counter increments is $(0.6, 0.2, 0.1, 0.1)$: for each counter increment, it has probability of $0.6$ happening in Player 1, $0.2$ in Player 2, $0.1$ in Player 3, and $0.1$ in Player 4. 

 
The knowledge on the probability distribution (i.e., the Multinomial Distribution) of the counter increments indeed can be leveraged to significantly reduce the communication cost.
As an extreme case in our earlier example in Figure~\ref{fig:example}, suppose that one knows the \emph{final} counter value $c_i$ of the $i$-th player for all $i\in \{1, 2, \ldots, k\}$ at the moment when their sum reaches $N$, namely, $c_1 = 30$, $c_2 = 10$, $c_3 = c_4 = 5$. 
A better solution is to instruct
the coordinator to send a \emph{customized} slack $s_i  = c_i$ to the $i$-th player, and to raise the alarm when it receives the fourth messages from the players. Clearly, the communication cost of this solution is only $8$ messages, \emph{five times} less than the cost of the $\cmy$ algorithm.     
While knowing all the final counter values $c_i$ a prior is, of course, too good to be true, this observation sheds a light on the possibility of designing improved \emph{data-dependent} algorithms with the knowledge of the counter increment distribution.


Motivated by the above, 
in this paper, we consider the DT problem under the assumption below:
\begin{assumption}
	The counter increments follow a certain Multinomial Distribution which can be unknown. More specifically, each counter increment occurs in the $i$-th player with probability $\mu_i$ for all $i \in \{1, 2, \ldots, k\}$, where $\sum_{i=1}^k \mu_i = 1$. 
\end{assumption}

\begin{table}[t]
\hspace{-1.5mm}
\begin{tabular}{|l|l|}
\hline
\textbf{Notation} & \textbf{Description}                              \\ \hline
$N$          & the threshold to monitor              \\ \hline
$k$          & the number of players                    \\ \hline
$[k]$        & the set of integers from $1$ to $k$      \\ \hline
$n_i$        & the counter of the $i$-th player         \\ \hline
$s_i$        & the slack of the $i$-th player         \\ \hline
$\mu_i$      & the probability that a new item arrives at player $i$ \\ \hline 
$\bar \mu_i$ & the estimation of $\mu_i$ \\ \hline 
$T(N)$       & the communication cost with threshold $N$ \\ \hline
\end{tabular}
\caption{Frequently used notations}
    \vspace{-7mm}
\end{table}

\vspace{1mm}
\noindent
{\bf Our contributions.} 
We make the following contributions:
\begin{itemize}[leftmargin = *]
	\item First, for the case that the \emph{concrete} Multinomial Distribution (i.e., the concrete values of all $\mu_i$'s) is \emph{known} a prior, we show a data-dependent algorithm, called $\DstT$, for solving the DT problem.
		The communication cost of the $\DstT$ algorithm is bounded by $O(k \log \log N)$ with high probability, improving the state-of-the-art $O(k \log \frac{N}{k})$ bound.
		We further propose the $\DstDT$ algorithm which improves the practical performance of $\DstT$, while retaining exactly the same communication bound.
	\item Second, for the case that the distribution is \emph{unknown}, we propose two learning based algorithms, called $\LnDstT$ and $\LnDstDT$, corresponding to the two algorithms in the first case. Both of these two algorithms can learn the data distribution on the fly, meanwhile achieving exactly the same theoretical bounds as their counterpart algorithms. 
	\item Moreover, we design an effective heuristics to optimize our algorithm implementations.
	\item Finally, we conduct extensive experiments on both three real datasets and four synthetic datasets of different data distributions. The experimental results show that our proposed algorithms outperform the state-of-the-art algorithms by consuming up to five times (i.e., 5x) less communication cost.
\end{itemize}

}

\section{Related Work}
The \emph{distributed tracking} (DT) problem has been well studied in terms of both upper bounds and lower bounds, since it was first proposed. 

Prior to the $\cmy$ algorithm, an uniform slack (called $\US$) algorithm was proposed by Cormode et al.~\cite{Cormode13}. The communication cost of $\US$ is bounded by $O(k^2 \log \frac{N}{k})$. 
The design of $\US$ is based on the observation that 
for $N$ items arriving at $k$ players, there must be at least one of player received $ N/ k$ items. 
The $\US$ algorithm runs in rounds. At the at the beginning of the first round, the coordinator broadcasts a slack $ N / k $ to each players. The player notifies the coordinator when its counter exceeds $N / k$. Upon receiving one notification, the coordinator informs the rest of players to report the values of their counters. This ends the first round. The number of items arrived, namely, $\sum_{i = 1}^k n_i$ is at least $ N / k $ and at most $ \lceil N / k \rceil + (k - 1) \lfloor N / k \rfloor \le N$. The coordinator updates the threshold $N \leftarrow N - \sum_{i = 1}^k n_i$. If $N = 0$, it raises an alarm, otherwise it starts a new round. Each round incurs $O(k)$ communications and decreases the threshold $N$ by a factor of at least $(1 - 1/ k) \le \exp({-1/k})$. Therefore, the number of rounds is at most $O(k \log \frac{N}{k})$ and the communication cost is thus  $O(k^2 \log \frac{N}{k})$. 

Later on, the $\cmy$ algorithm was proposed in \cite{CMY11}. As introduced earlier, the $\cmy$ algorithm consumes $O(k\log \frac{N}{k})$ communication cost. Although we are focusing on the number of messages in this paper, in~\cite{CMY11}, Cormode et al. showed a $\Omega(k \log \frac{N}{k})$-bit communication lower bound for the DT problem. Moreover, the $\cmy$ algorithm indeed admits a bit-version implementation with communication cost of $O(k \log N)$ bits.

More works have been done for the variants of the DT problem. Randomized algorithm was also proposed for approximate count tracking \cite{CMY11}. Instead of reporting exactly the $N$-th item, the coordinator is allowed to raise an alarm on the arrival of any item between $[(1 - \epsilon)N, N]$, where $\epsilon \in (0, 1)$ is a specified parameter. The problem is easier than the exact count tracking problem due to the relaxation and the proposed algorithm has cost $O(\frac{1}{\epsilon^2} \log \frac{1}{\delta})$ bits \cite{CMY11}, where $\delta$ is the failure probability of the algorithm. \cite{KCR06, HYZ19} studied the continuous count tracking problem, in which the coordinator is required to report an estimation $\hat n$ of $n = \sum_{i \in [k]} n_i$ at any time stamp, such that $\hat n \in (1 \pm \epsilon) n$. Their algorithms have communication costs of $O(\frac{k}{\epsilon} \log \frac{\epsilon N}{k})$ \cite{KCR06} and $O(\frac{\sqrt k}{\epsilon} \log N)$ messages \cite{HYZ19}, respectively. The work of \cite{CMY11} also considered threshold tracking of $F_p$ moment, where DT can be considered as a special case of $p = 1$.


Data pattern has been studies and exploited for different distributed monitoring queries (\cite{CGMR05, GDGSS12}), but none of them targets the fundamental threshold count tracking problem. 
\cite{CGMR05} introduced the Update-Rate Model for distributed tracking of approximate quantiles, which assumes that items arrives at the $i$-th player at a local rate specified to $i$.  Note that this is captured by the multinomial distribution model if we normalize the rates by the summation of the players' rates (and with proper scaling of time). Later 
\cite{GDGSS12} extends the idea for geometric monitoring to reduce the communication cost. 

Besides, interestingly, the techniques for solving the DT problem has also been applied to solve some seemly ``remote'' problem in the \emph{single machine} setting~\cite{QGT16}.

\begin{algorithm}[t]
	\caption{The $\ell$-Notifications-to-End Framework}\label{algo:framework}
	\begin{flushleft}
		\hspace*{\algorithmicindent} \textbf{Input:} a threshold $N$, the number of players $k$, and \\
		\hspace*{\algorithmicindent} \hspace*{\algorithmicindent} \hspace*{\algorithmicindent} possibly a failure probability $\delta$ if applicable \\
    \end{flushleft}
	\begin{algorithmic}[1]
	    \State{Set $n_i \leftarrow 0$ for all $\forall i \in [k]$} \label{step:dt:initialization}
	    \State{If $N \leq 4k$, run the straightforward algorithm until it terminates.}
	    \State{If $N  \leq  \beta \cdot k \ln\frac{k}{\delta}$, run the $\cmy$ algorithm until it terminates, where $\beta$ is an algorithm-specified constant.} 
	    	\State{The coordinator \emph{\bf send a slack $s_i$} to player $i$ for $\forall i \in [k]$.}
		\State{Each Player $i$ notifies the coordinator, \emph{\bf when $n_i$ meets certain condition with respect to $s_i$}.}
		\State{The coordinator collects all the $n_i$'s from the players,  \emph{\bf when it receives the $\ell$-th notification}.}
		\State{Set $N \leftarrow N - \sum_{i = 1}^k n_i$}
		\State{If $N = 0$, the coordinator raises an alarm and terminate.}
		\State{Otherwise, go to Step~\ref{step:dt:initialization} and start a new round.}
	\end{algorithmic}
\end{algorithm}

\section{A Unified Algorithm Framework}
Before we get into the details of our algorithms, in this section, we first propose a \emph{unified algorithm framework}, called \emph{$\ell$-Notifications-to-End}, where $\ell$ is a \emph{characteristic parameter} of an algorithm and not an input parameter. 
As we will see shortly, all our algorithms, the $\cmy$ algorithm as well as the $\US$ algorithm are all under this framework. 
Algorithm~\ref{algo:framework} shows the pseudo code of the framework.

\vspace{1mm}
\noindent
{\bf Algorithm characteristics.} Essentially, algorithms under the \emph{$\ell$-Notifications-to-End} framework only differ in the following three \emph{characteristics}:
\begin{itemize}[leftmargin = *]
	\item {\bf Characteristic 1:} the value of slack $s_i$ for $\forall i\in [k]$ (Line 4);
	\item {\bf Characteristic 2:} the condition for a player to notify the coordinator (Line 5);
	\item {\bf Characteristic 3:} the number $\ell$ of notifications received to end a round (Line 6).
\end{itemize}
To see this, consider the $\cmy$ algorithm, where: (i) $s_i = \lf \frac{N}{2k} \rf$; (ii) a player notifies the coordinator when $n_i$ is increased by $s_i$; and (iii) $\ell = k$. Thus, the $\cmy$ algorithm is an $k$-Notifications-to-End algorithm.   
On the other hand, the $\US$ algorithm is, in fact, a $1$-Notification-to-End algorithm with $s_i = N/k$, where a player notifies the coordinator when $n_i \geq s_i$, and with Line 3 being never executed. 
As we will see in the next two sections, all our algorithms will be focusing on designing the above three characteristics.

\vspace{1mm}
\noindent
{\bf Communication cost expression.} Observe that the communication cost for executing Line 2 and Line 3 in Algorithm~\ref{algo:framework} are bounded by $O(k)$ and  $O(k\log \log \frac{k}{\delta})$, respectively.
As these two cases are easy to check and solve, it suffices to focus on the case $N = \omega(k \ln \frac{k}{\delta})$. 
In this case, it can be verified that an $\ell$-Notification-to-End algorithm consumes $O(k + \ell)$ communication cost per round. 
Therefore, the overall communication cost in this case is bounded by $O((k + \ell)\cdot R + k \log\log \frac{k}{\delta})$, where $R$ is the total number of rounds that have been executed before entering into Line 2.

\section{Tracking with Known Distribution}
{\newblue In this section, we consider the case that the concrete Multinomial Distribution of the counter increments (a.k.a. the item arrivals) is known. That is, the concrete values of $\mu_i$ for $i\in [k]$ are given.
This case allows us to just focus on algorithmic design without worrying too much about the learning of the distribution.
}


\subsection{Tracking with Static Slacks}

{\newblue

\noindent
{\bf The challenges.}
We note that even the concrete values of $\mu_i$'s are known in advance, the problem is still challenging. 

As the concrete values of all $\mu_i$'s are known, it is natural to think about modifying the $\cmy$ algorithm such that 
the slack $s_i$ is set to the expected number of items that Player $i$ will receive when the $N$-th item arrives, namely, $s_i = \mu_i N$ for $\forall i \in [k]$. 
We denote this new $k$-Notifications-to-End algorithm by $\mathcal{A}$. 
 
Unfortunately, in general, it is unlikely that every player will receive \emph{exactly} $\mu_i N$ items, i.e., $n_i = \mu_i N$,  when the $N$-th item arrives; the actual $n_i$ could be more or less than $\mu_i N$. As a result, $\mathcal{A}$ may fail to capture the moment of the $N$-th item's arrival, when the coordinator receives the $k$-th notification.   

To remedy this, one possible way is to further modify $\mathcal{A}$ into a $1$-Notifications-to-End algorithm by setting $\ell$ to 1, where the coordinator in $\mathcal{A}$ collects all the $n_i$'s as soon as it receives a notification from the player. 
This guarantees that 
$\sum_{i = 1}^k n_i < N$ and the coordinator would not miss the $N$-th item.  
However, $\mathcal{A}$ could be less efficient than $\cmy$. This is because any small deviation from $\mu_i N$ would easily make $\mathcal{A}$ end the round too early, resulting in an increase on the number of rounds $R$. 
As an illustration, suppose that there are $10$ players and the first three players have very low probabilities of receiving items. In particular, $\mu_1 = \mu_2 = \mu_3 = \frac{1}{N}$. In this case, if any item arrives at any of the first three players, a notification is sent to the coordinator and the round ends. With probability at least $1 - 1/e$, this round captures only $N / 3$ items: for a given item, the probability that it arrives at a player other than the first three is $(1 - 3 / N)$; therefore, with probability $(1 - 3 / N)^{N / 3} \le 1 / e$, none of the first $N / 3$ items arrive at any of the first three player. In comparison, $\CMY$ captures at least $N/2$ in one round. 


To capture more items in one round, the slack should contain a leeway for the player's counter to bear deviations from its expectation, i.e., $s_i$ should be greater than $\mu_i N$ for $i \in [k]$. 
The player should receive a few more items than its expectation to defer the notification. However, this conflicts with our goal of capturing the $N$-th item -- if we set $s_i > \mu_i N$ for $i \in [k]$, then $\sum_{i \in [k]} s_i > N$. The coordinator could miss the arrival of the $N$-th item. 
The following constraint for ensuring the correctness 
\begin{equation} \label{ineq:correctness constraint}
    \sum_{i = 1}^k s_i \le N
\end{equation}
implies that 
tracking $N$ items in just one round seems too ambitious. 
We relax our goal and, instead, aim to track $t$ items for some $t < N$ and set $s_i > \mu_i t$, while ensuring the correctness, i.e., $\sum_{i=1}^k s_i \leq N$. 
We prove that there exists $s_i$'s, such that when the first $t$ items arrives, with high probability that none of the player has received more than $s_i$ items. It implies that no notification would have been sent to the coordinator. 
In other words, when the first notification is sent, more than $t$ items have arrived in this round. 
Clearly, the larger $t$ is, the less number of rounds we need.
}

\vspace{1mm}
\noindent
{\bf The $\DstT$ Algorithm.}
Motivated by the above observation, we propose our first data-dependent algorithm, $\DstT$, which is an $1$-Notifications-to-End algorithm. 
characterized by the followings:
\begin{itemize}[leftmargin = *]
	\item {\bf Characteristic 1}: 
		\begin{equation}\label{eq:si}
		s_i = \mu_i t + \sqrt{2 t \mu_i (1 - \mu_i) \ln \frac{k}{\delta} } + \frac{2}{3} \ln \frac{k}{\delta}, \forall i \in [k];
		\end{equation}
	\item {\bf Characteristic 2}: Player $i$  notifies the coordinator when $n_i = s_i$;
	\item {\bf Characteristic 3}: $\ell = 1$, that is, the coordinator ends a round when it receives the first notification from the players.
\end{itemize}
Substituting the above implementations of the three characteristics to Lines 4, 5 and 6, respectively, in Algorithm~\ref{algo:framework} gives the pseudo code of the $\DstT$ algorithm.
Furthermore, we have the following key theorem:

\begin{theorem} \label{thm:cost_of_known_dt}
With probability $1 - \delta'$, the $\DstT$ algorithm:
\begin{itemize}[leftmargin = *]
	\item runs at most $O(\log \log \frac{N}{k \log \frac{k}{\delta}} + \log \log \frac{k}{\delta})$ rounds;
	\item has total communication cost $O(k \log \log \frac{N}{k \log \frac{k}{\delta}} + k \log \log \frac{k}{\delta})$,
	\end{itemize}
	where $\delta' = \delta \cdot O(\log \log N)$.
\end{theorem}

Answering the following three questions is the key to proving Theorem~\ref{thm:cost_of_known_dt}:
\begin{itemize}[leftmargin = *]
\item Why does Expression~\eqref{eq:si} ensure that $\DstT$ can capture at least $t$ items at one round with probability $1- \delta$? 
\item How large can $t$ be? 
\item How many rounds does $\DstT$ need? 
\end{itemize}
In what's follows, we address these questions one by one. 

\vspace{1mm}
\noindent \textbf{Setting the Slack.} Expression~\eqref{eq:si} is actually derived from the following concentration inequality: 
\begin{fact}
\label{lma:Bernstein inequality} (Bernstein inequality \cite{CL06}) 
Let $Y_1, \ldots, Y_t$ be independent, random variables. Let $Y = \sum_{j = 1}^t Y_j$, and $M > 0$ be such that $Y_j \le E[Y_j] + M$ for all $j \in [t]$. For any $\lambda \ge 0$, 
\begin{equation}
    \Pr[Y \ge E[Y] + \lambda] \le \exp{\left( - \frac{\lambda^2}{2(Var[Y] + M \lambda / 3) } \right). }
\end{equation} 
\end{fact} 
\vspace{2mm}

Consider a fixed Player $i$ and the first $t$ items arrived in the current round.
Denote by $X_j$ the Bernoulli random variable that $X_j = 1$ if the $j$-th item arrives at player $i$, and $X_j = 0$ otherwise. Then $E[X_j] = \mu_i$, $Var[X_j] = \mu_i (1- \mu_i)$ and $X_j \le E[X_j] + 1$ for $j \in [t]$. By dentition, the counter $n_i = \sum_{j \in [t]} X_j$. Fact~\ref{lma:Bernstein inequality} gives the following results:

\begin{lemma}
     \label{lma:setting UB}
    Given $t > 0$ and a failure probability $\delta > 0$, define \begin{equation} \label{eqa:setting UB}
        UB_i = \mu_i t + \sqrt{2 t \mu_i (1 - \mu_i) \ln \frac{k}{\delta} } + \frac{2}{3} \ln \frac{k}{\delta}.
    \end{equation} 
    Then the probability $Pr[n_i \geq UB_i] \leq \frac{\delta}{k}$. Likewise, for 
    \begin{equation} \label{eqa:setting LB}
        LB_i = \mu_i t - \sqrt{2 t \mu_i (1 - \mu_i) \ln \frac{k}{\delta} } - \frac{2}{3} \ln \frac{k}{\delta},
    \end{equation} 
    we have $Pr[n_i \le LB_i] \leq \frac{\delta}{k}$. 
\end{lemma}

The proof of Lemma~\ref{lma:setting UB} can be found in the Appendix~\ref{APPENDIX_PROOFS}. 
By Lemma~\ref{lma:setting UB}, for a fixed $i$, by setting the slack $s_i = UB_i$, when the first $t$ items arrive, the event $n_i \geq s_i$ happens with probability at most $\delta /k $. 
By union bound, the probability that $n_i \ge s_i$ for any $i \in [k]$ is at most $\delta$. 
Therefore, the following corollary holds. 
\begin{corollary} \label{cor:capture_t}
With probability $\geq 1 - \delta$, 
the coordinator receives no notification from the players for the first $t$ items in a round. 
\end{corollary} 

\vspace{1mm}
\noindent 
\textbf{Setting $t$.} Define function $f(t) \doteq \sum_{i\in [k]} s_i$, with $s_i = UB_i$ (Expression~\eqref{eq:si}). It is easy to verify that $f(t)$ is monotonically increasing with $t$. Clearly, the larger $t$ the more items can be tracked in a round, while $t$ should also satisfy: $f(t) \leq N$ to ensure the correctness of the algorithm. 
As a result, it is desired to maximise $t$ subject to $f(t)\leq N$. 
This will possibly reduce the total number of rounds in the $\DstT$ algorithm and hence, reduce the total communication cost.
However, computing the optimal $t$ precisely may not be an easy task.
Nonetheless, as we show shortly, $t =  N - (\sqrt{2} + 2/3 ) \sqrt{k N \ln \frac{k}{\delta} }$ is already good enough for our purpose. 
In particular, we have the following lemma whose proof is in given in Appendix~\ref{APPENDIX_PROOFS}.
\begin{lemma}
     \label{lma:setting t}
     $f(t) \leq N$ for $t = N - (\sqrt{2} + 2/3 ) \sqrt{k N \ln \frac{k}{\delta} }$.
\end{lemma} 

\vspace{1mm}
\noindent \textbf{Bounding the communication cost.}
Consider an implementation of the $\DstT$ algorithm with its three characteristics plug in to the unified framework (Algorithm~\ref{algo:framework}), where we further explicitly set 
$\beta = 2 \cdot (\sqrt{2} + 2/3)^2$.  
According to this implementation,
we know that when
$N \ge \beta \cdot {k \ln \frac{k}{\delta}}$, 
at the end of each round, the value of $N$ can be decreased to 
at most $(\sqrt{2} +2/3) \sqrt{k N \ln \frac{k}{\delta} }$ with probability $1 - \delta$ (by Corollary~\ref{cor:capture_t} and Lemma~\ref{lma:setting t}).
Furthermore, when $N$ is found smaller than $\beta \cdot {k \ln \frac{k}{\delta}}$ at the start of a round, the algorithm switches to the $\cmy$ algorithm (according to Line 3 in Algorithm~\ref{algo:framework}). Therefore, this gives the following recursion, where $T(N)$ denotes the communication cost of the algorithm with respect to $N$.
\begin{align*}
    T(N) = \begin{cases}
        T \left( (\sqrt{2} +2/3) \sqrt{k N \ln \frac{k}{\delta} } \right) + O(k), 
        & \hspace{-1mm} N \ge 2 (\sqrt{2} +2/3)^2 k \ln \frac{k}{\delta} \\
        T(N / 2) + O(k),  
        & \hspace{-8mm} 4k \le N < 2 (\sqrt{2} + 2/3)^2 k \ln \frac{k}{\delta} \\
        O(k),   
        & \hspace{-1mm} N < 4k
    \end{cases}
\end{align*}
Solving the recursion gives the last lemma we need for Theorem~\ref{thm:cost_of_known_dt}.
\begin{lemma} \label{lma:recursion_of_known_dt}
    $T(N) = O(k \log \log \frac{N}{k \ln \frac{k}{\delta}} + k \log \log \frac{k}{\delta} ).$ 
\end{lemma}

\vspace{1mm}
\noindent
{\bf Proving Theorem~\ref{thm:cost_of_known_dt}.}
Putting Corollary~\ref{cor:capture_t}, Lemma~\ref{lma:setting t} and ~\ref{lma:recursion_of_known_dt} together, it thus completes the proof for Theorem~\ref{thm:cost_of_known_dt}.

\subsection{Tracking with Dynamic Slacks}

In the previous subsection, we know that the $\DstT$ algorithm assigns to Player $i$ a slack $s_i = UB_i > \mu_i t$. In expectation, $\mu_i t$ items arrives at the $i$-th player. 
Intuitively, $UB_i - \mu_i t$ is the \emph{tolerance} that how much the counter $n_i$ is allowed to deviate from its expectation $\mu_i t$, when $t$ items arrives. As soon as $n_i$ reaches $UB_i$, Player $i$ is not allowed to further receive any new items (to ensure the correctness). Hence, at this moment, the coordinator collects the precise counters and ends the current round. 

While the above strategy has been shown to be effective in the previous subsection, setting 
$s_i = UB_i$ is a \textit{static} slack assignment strategy. 
In the sense that,  the tolerance for deviations, i.e., $UB_i - \mu_i t$,  is \emph{pre-determined} and fixed for each Player $i$.
When the coordinator ends a round, except for the player sending the notification, all other players actually have not fully used up their deviation tolerance.
An immediate question comes up: Can we further improve the utilization of those \emph{non-fully-used} deviation tolerances, before ending a round? 

Motivated by the question, we design a new slack assignment strategy to \emph{dynamically} adjust the deviation tolerance for the players. The basic idea is as follows. First, observe that the sum of all the deviation tolerance of the players is computed as
$$\sum_{i = 1}^k (UB_i - \mu_i t) \leq N - t.$$
Instead of pre-assigning a static tolerance to each player, we adopt the strategy of the $\cmy$ algorithm. More specifically, the coordinator sends a \emph{base} value $b_i = \mu_i t$, and a deviation tolerance $s_i = \lf \frac{N - t}{2k} \rf$ to Player $i$, for $\forall i \in [k]$.
A player sends a notification to the coordinator for every counter increment $s_i$ \emph{only when} $n_i \geq b_i$. 
The coordinator collects the counters and ends the round when it receives the $k$-th notification.
The resulted algorithm is called $\DstDT$; since a round is ended when the coordinator receives $k$ notifications, the $\DstDT$ algorithm is a $k$-Notifications-to-End algorithm, according to our unified framework. 

In particular, $\DstDT$ implements the three characteristics as follows:
\begin{itemize}[leftmargin = *]
	\item {\bf Characteristic 1:} the slack is a pair $(b_i, s_i)$ for $\forall i\in [k]$, where $b_i = \mu_i t$ and $s_i = \lf \frac{N-t}{2k} \rf$;
	\item {\bf Characteristic 2:} Player $i$ sends a notification to the coordinator for every counter increment $s_i$ \emph{only when} $n_i \geq b_i$;
	\item {\bf Characteristic 3:} $\ell = k$. 
\end{itemize}
Substituting the above implementations to the algorithm framework (Algorithm~\ref{algo:framework}), gives the pseudo code of the $\DstDT$ algorithm.

As strategy of $\cmy$ can guarantee that the coordinator will not miss the arrival of the $(N-t)$-th item, it thus guarantees that no more than  $N$ items can be received in a round. Therefore, the correctness of the $\DstDT$ algorithm follows.
Furthermore, the theorem below shows that with probability at least $1 - \delta$, at the end of a round, at least $t$ items arrive to the players.
\begin{theorem} \label{thm:correctness_of_dynamic_known_dt}
With probability at least $1 - \delta$, less than $k$ notifications will be sent from the players for the first $t$ items in a round.
\end{theorem} 

By Theorem~\ref{thm:correctness_of_dynamic_known_dt}, Lemmas~\ref{lma:setting t} and~\ref{lma:recursion_of_known_dt}, we have:

\begin{theorem}
	The $\DstDT$ algorithm achieves exactly the same bounds of $\DstT$ as stated in Theorem~\ref{thm:cost_of_known_dt}.
\end{theorem}

\section{Learning Based Tracking}

In this section, we consider the case that the underlying counter increment distribution is \emph{unknown}.
The basic idea is to learn the distribution on the fly. 
To learn the unknown distribution, we run the $\cmy$ algorithm for the \emph{first round}. 
This allows us to receive at least $N/2$ items with only $O(k)$ communication.
At the end of this first round, we estimate $\mu_i$ by $\bar \mu_i = \frac{n_i}{\sum_{i\in [k]} n_i}$ for $\forall i \in [k]$.
These $\bar \mu_i$'s are used in the subsequent rounds to determine the slacks. 
As $\bar \mu_i$'s are just estimations, they may introduce additional errors. 
Furthermore, since $\bar \mu_i$ could be an underestimation of $\mu_i$, the upper bound $UB_i$ computed by simply replacing $\mu_i$ with $\bar \mu_i$ in Expression~\eqref{eqa:setting UB} may be no longer a proper upper bound for $n_i$ when the first $t$ items arrive. 
Therefore, modifications to the previous algorithms are required. 

The modifications consist of three steps. First we construct some $\hat \mu_i$ based on $\bar \mu_i$ such that it is guaranteed that $\hat \mu_i \ge \mu_i$. Next, we show how to construct $UB_i$ with $\hat \mu_i$. Last, $t$ needs to be change to ensure $\sum_{i \in [k]} UB_i \le N$.

\vspace{1mm}
\noindent \textbf{Upper bound on $\mu_i$.} 
The concentration inequality below is needed.
\begin{fact} (Empirical Bernstein Bound) \cite{AMS09} \label{thm:empirical bernstein bound}
Let $Y_1, ..., Y_w$ be independent, random variables with mean $\mu$. Let $\bar Y =  \frac{1}{w} \sum_{j \in [w] } Y_j$, and $M > 0$ be such that $|Y_j| \le M$ for all $j \in [w]$. With probability at most $\delta$, it holds that 
\begin{equation*}
    |\bar Y - \mu | \ge  \sqrt{\frac{2\bar \sigma^2 \ln \frac{3}{\delta} }{w}  } + \frac{3M \ln \frac{3}{\delta} }{w} 
\end{equation*}
where  $\bar \sigma^2$ is the empirical variance of $Y_j$'s: 
$\bar \sigma^2 = 1/w \sum_{j \in [w]} (Y_j - \bar Y)^2$. 
\vspace{-2mm}
\end{fact}

Consider a fixed $i\in [k]$ and the number of items $n_i$ that arrive at Player $i$, for the $w$ items tracked in the first round. 
Denote $Y_j$ the Bernoulli random variable such that $Y_j = 1$ if the $j$-th item arrives at Player $i$, and $Y_j = 0$ otherwise. Then $Y_j \le 1$ for $j \in [w]$. Denote $\bar \mu_i = \bar Y = 1 / w \sum_{j \in [w]} Y_j$ as the empirical mean. 
As $Y_j$'s are Bernoulli random variables, the empirical variance is $\bar \sigma^2 = \bar \mu_i (1 - \bar \mu_i)$. 
According to Fact~\ref{thm:empirical bernstein bound}, a upper bound $\hat \mu_i$ on $\mu_i$ can be obtained:
\begin{equation}
    \label{eq:learned_upper}
    \hat \mu_i \doteq \bar \mu_i + \sqrt{\frac{2 (\bar \mu_i - (\bar \mu_i)^2)  \ln \frac{3}{\delta} }{w}  } + \frac{3 \ln \frac{3}{\delta} }{w}
\end{equation}

\noindent \textbf{Modification on $UB_i$.}
The slack $s_i = UB_i$ is modified as below, 
\begin{equation} \label{eqa:cold start upper}
    UB_i = \hat \mu_i t + \sqrt{2 t \hat \mu_i \ln \frac{k}{\delta} } + \frac{2}{3} \ln \frac{k}{\delta}
\end{equation}

\noindent \textbf{Modification on $t$.} We need to change the value of $t$. Define $\hat \Sigma = \sum_{i \in [k]} \hat \mu_i$. Then we set
\begin{equation} \label{eqa:cold start t}
   t = N / {\hat \Sigma} - (\sqrt{2} + \frac{2}{3} ) \sqrt{k (N / {\hat \Sigma}) \ln \frac{k}{\delta} }
\end{equation}

Substituting the modified $\hat u_i$, $UB_i$ and $t$ to the known-distribution counterparts, we can have the learning based versions for tracking with static slack (called $\LnDstT$) and with dynamic slacks (called $\LnDstDT$) respectively. Some extra care is required to set the constant $\beta = 2\cdot(2\sqrt{2} + 2/3 + 3)^2$ in the condition of when to switch to the $\cmy$ algorithm (at Line 3 in Algorithm~\ref{algo:framework}) in the framework. 
Moreover, we show our final theorem whose proof can be found in Appendix~\ref{APPENDIX_PROOFS}.
\begin{theorem} \label{thm:time_cold_start}
	With probability at least $1 - \delta$, the communication cost of the $\LnDstT$ algorithm (respectively, the $\LnDstDT$ algorithm) is bounded by $O(k \log \log \frac{N}{k \ln \frac{k}{\delta}} + k \log \log \frac{k}{\delta} ))$.
\end{theorem}

\section{Experimental Evaluation}
 
This section evaluates the proposed algorithms against the state-of-art competitors on a machine running on Ubuntu 18.04 with Intel(R) Core(TM) i7-8665U CPU @1.90 GHz and 16GB memory. We compare our four algorithms: $\DstT$, $\LnDstT$, $\DstDT$, $\LnDstDT$ with $\DT$ and $\US$. 
All the algorithms are implemented by C++ and compiled with gcc 7.4.0. A backup heuristic is implemented such that, when the empirical distribution is not stable, our algorithms can detect this case and switch to $\cmy$. 
The details can be found in Appendix~\ref{appendix:opt}. 
We conduct experiments on three real datasets and four synthetic datasets; the meta data are summarised in Table~\ref{tab:datasets}.

\begin{table}[t]
\begin{tabular}{|c|c|c|}
\hline
\hspace{-2mm} \bf{\emph{Name}}                              & \hspace{-2mm} {\bf \emph{Threshold}} ($N$)              & {\bf \emph{\#Players}} ($k$)         \\ \hline\hline
\hspace{-2mm} WorldCup Day 30/60/90             & \hspace{-2mm} 3.4 M/48 M/1.8 M    & \hspace{-2mm} 8/29/2      \\ \hline
\hspace{-2mm} Dartmouth 1st Oct/Nov/Dec         & \hspace{-2mm} 184 K/254 K/297 K & \hspace{-2mm} 336/348/319 \\ \hline
\hspace{-2mm} Uber Feb/Apr/June                 & \hspace{-2mm} 2.2 M/2.2 M/2.8 M & \hspace{-2mm} 262/262/262 \\ \hline
\hspace{-2mm} Uniform                           & \hspace{-2mm} $2^{10} - 2^{24}$ & \hspace{-2mm} $2 - 256$       \\ \hline
\hspace{-2mm} Gaussian                          & \hspace{-2mm} $2^{10} - 2^{24}$ & \hspace{-2mm} $2 - 256$                 \\ \hline
\hspace{-2mm} Zipfian                           & \hspace{-2mm} $2^{10} - 2^{24}$ & \hspace{-2mm} $2 - 256$                 \\ \hline
\hspace{-2mm} Exponential                       & \hspace{-2mm} $2^{10} - 2^{24}$ & \hspace{-2mm} $2 - 256$                 \\ \hline
\end{tabular}
\caption{Dataset Characteristics ($M = 10^6$ and $K = 10^3$)}
\label{tab:datasets}
\vspace{-5mm}
\end{table}

\begin{figure*}[ht!]
	 \vspace{-2mm}
	 \includegraphics[width=1.0\linewidth]{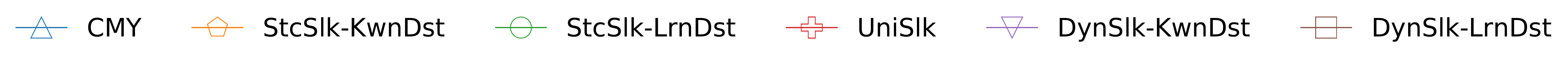} \\
    \resizebox{1.0\textwidth}{!}{%
	 \begin{tabular}{ccc} 
	\includegraphics[width=0.26\linewidth]{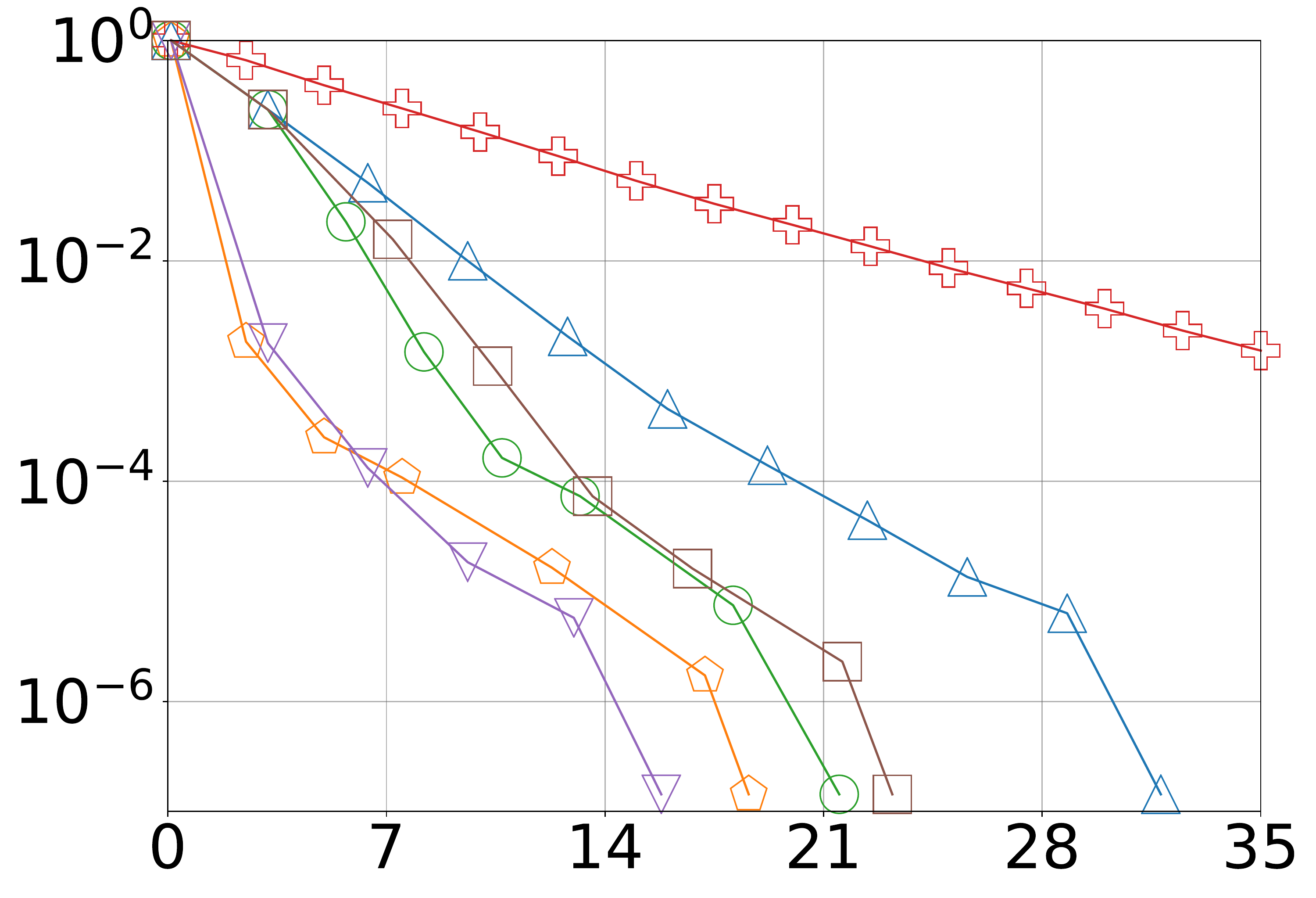} & 
	\includegraphics[width=0.26\linewidth]{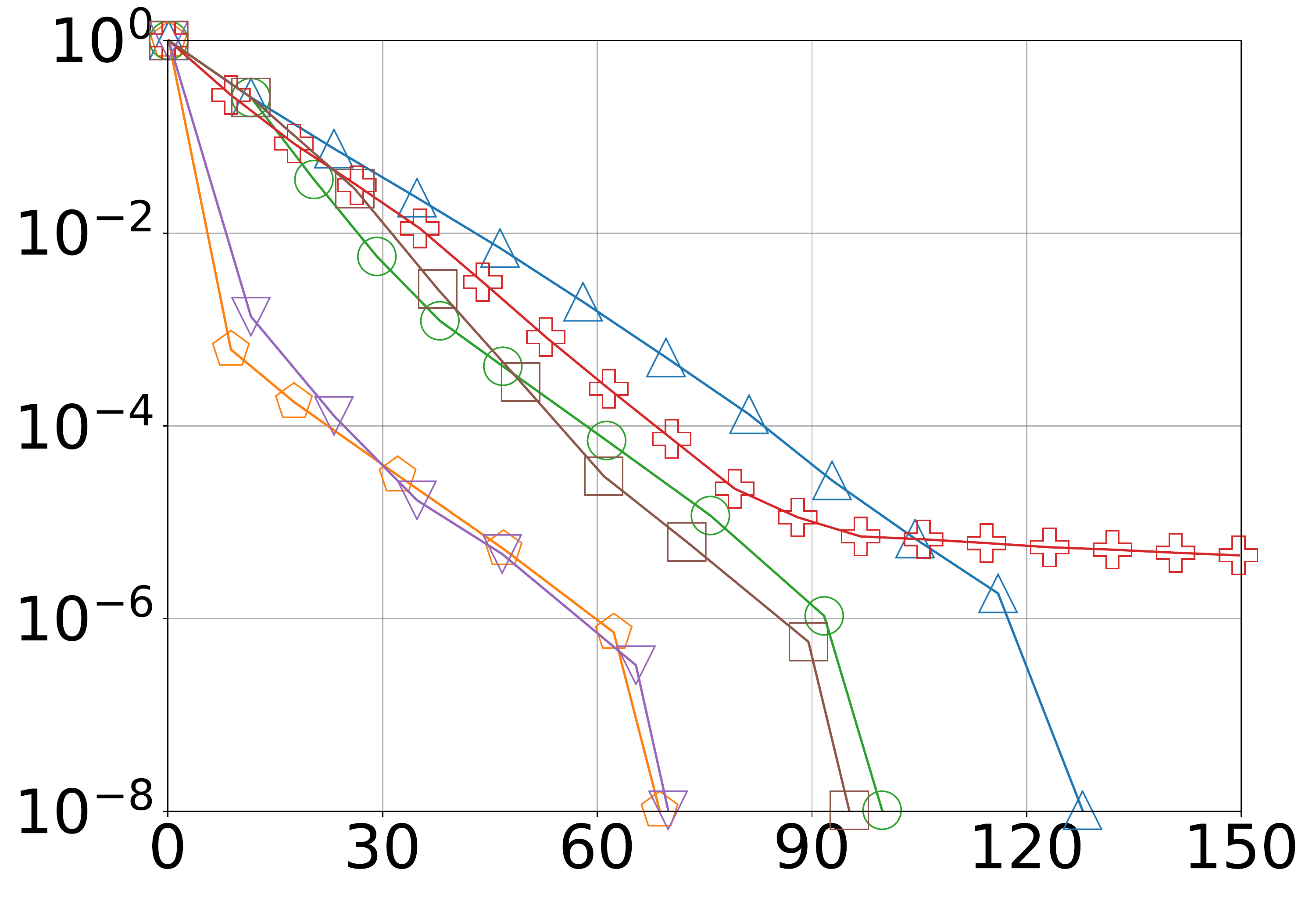} & 
	\includegraphics[width=0.26\linewidth]{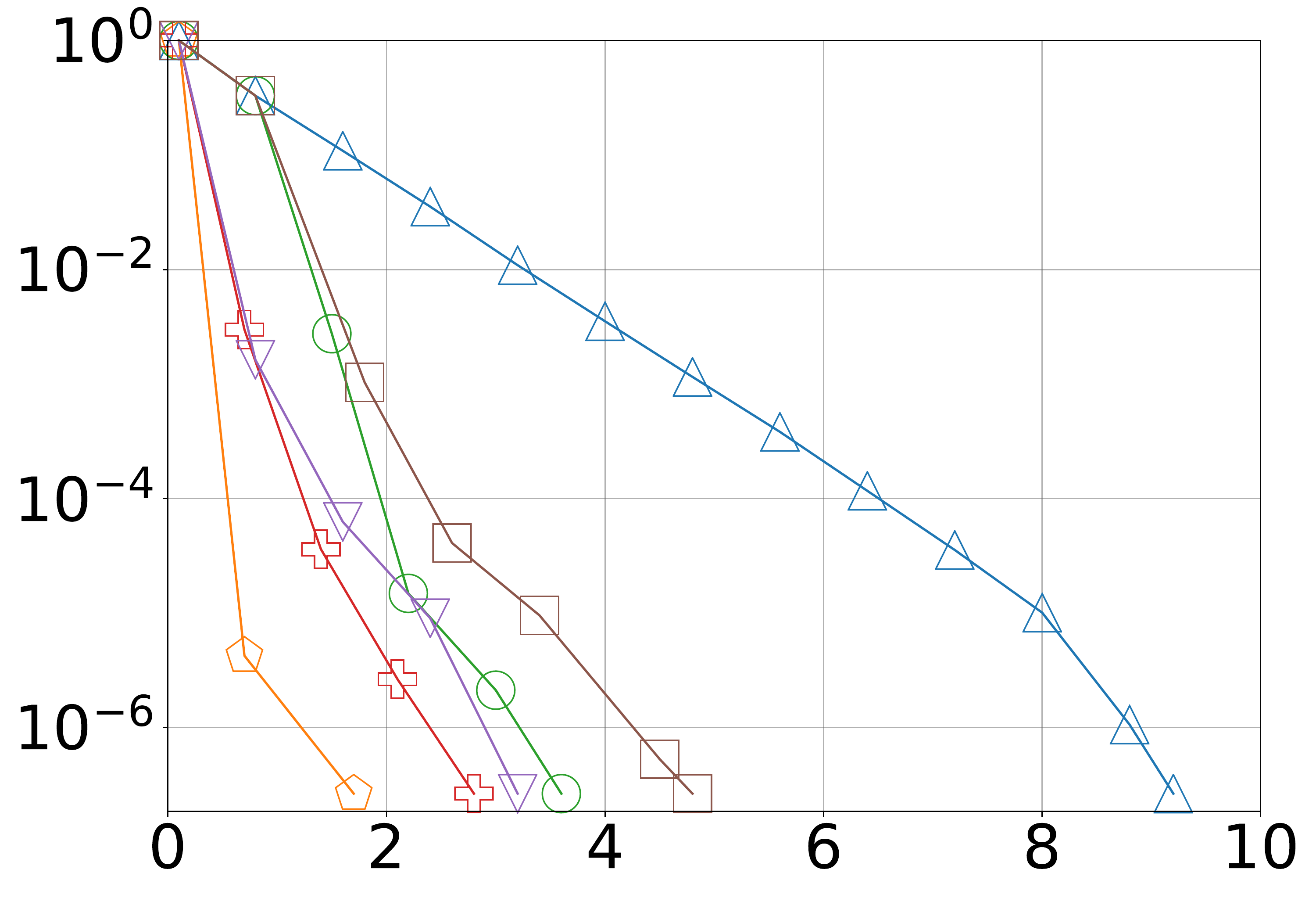} \\[-2mm]
	(a) {\em WorldCup Day 30} & 
	(b) {\em WorldCup Day 60} & 
	(c) {\em WorldCup Day 90}  \\[1mm]
	\end{tabular}
	}
	\vspace{-3mm}
     \figcapup 
     \caption{The Untracked Item Percentage v.s. Communication Cost ($10$x) on WorldCup datasets} 
     \label{fig:untracked_percent_vs_communication_real_WorldCup}
     \figcapdown
     \vspace{-1.5mm}
\end{figure*}

\begin{figure*}[ht!]
	 \vspace{-2mm}
 \centering
    \resizebox{1.0\textwidth}{!}{%
	 \begin{tabular}{ccc} 
	\includegraphics[width=0.26\linewidth]{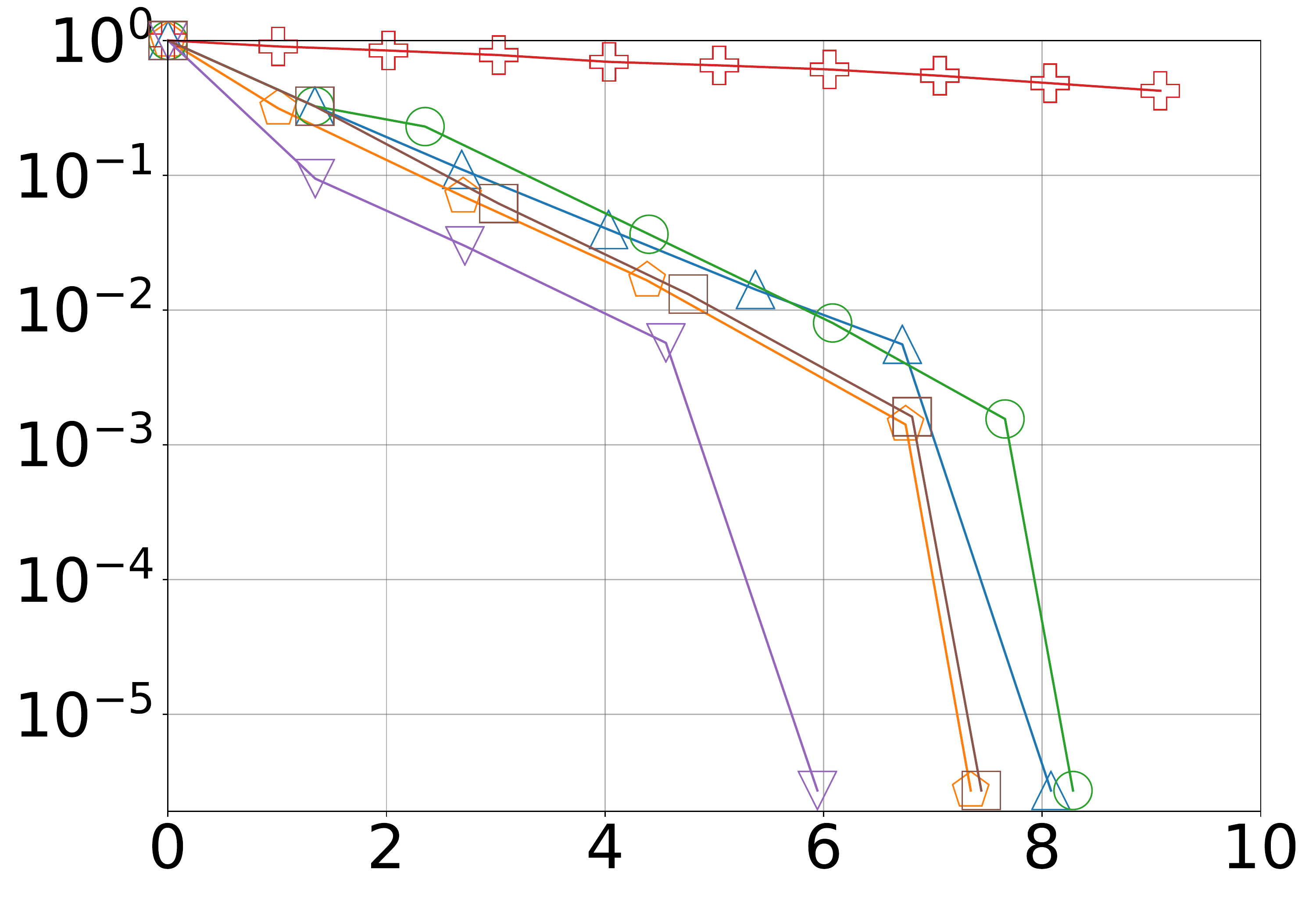} & 
	\includegraphics[width=0.26\linewidth]{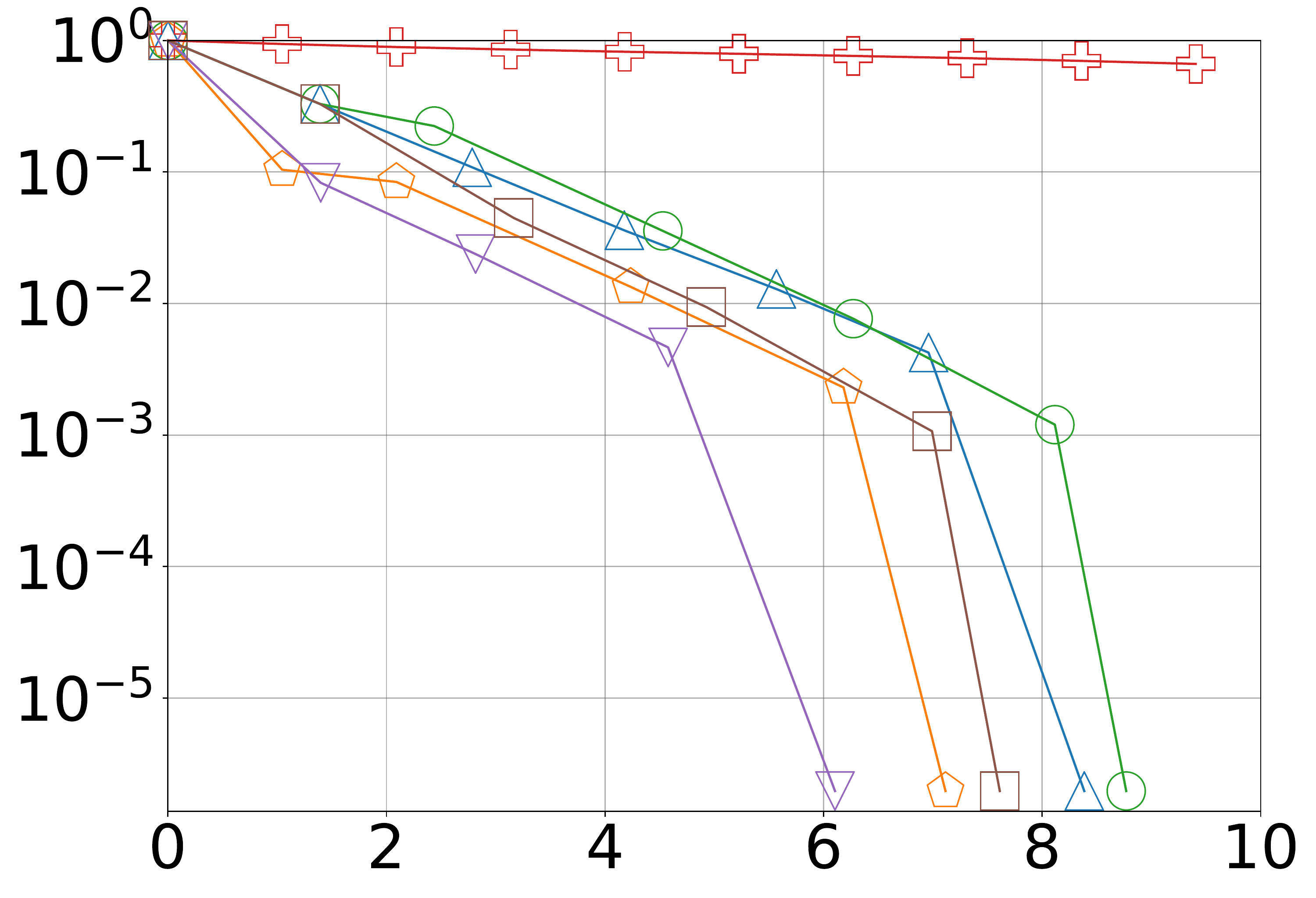} & 
	\includegraphics[width=0.26\linewidth]{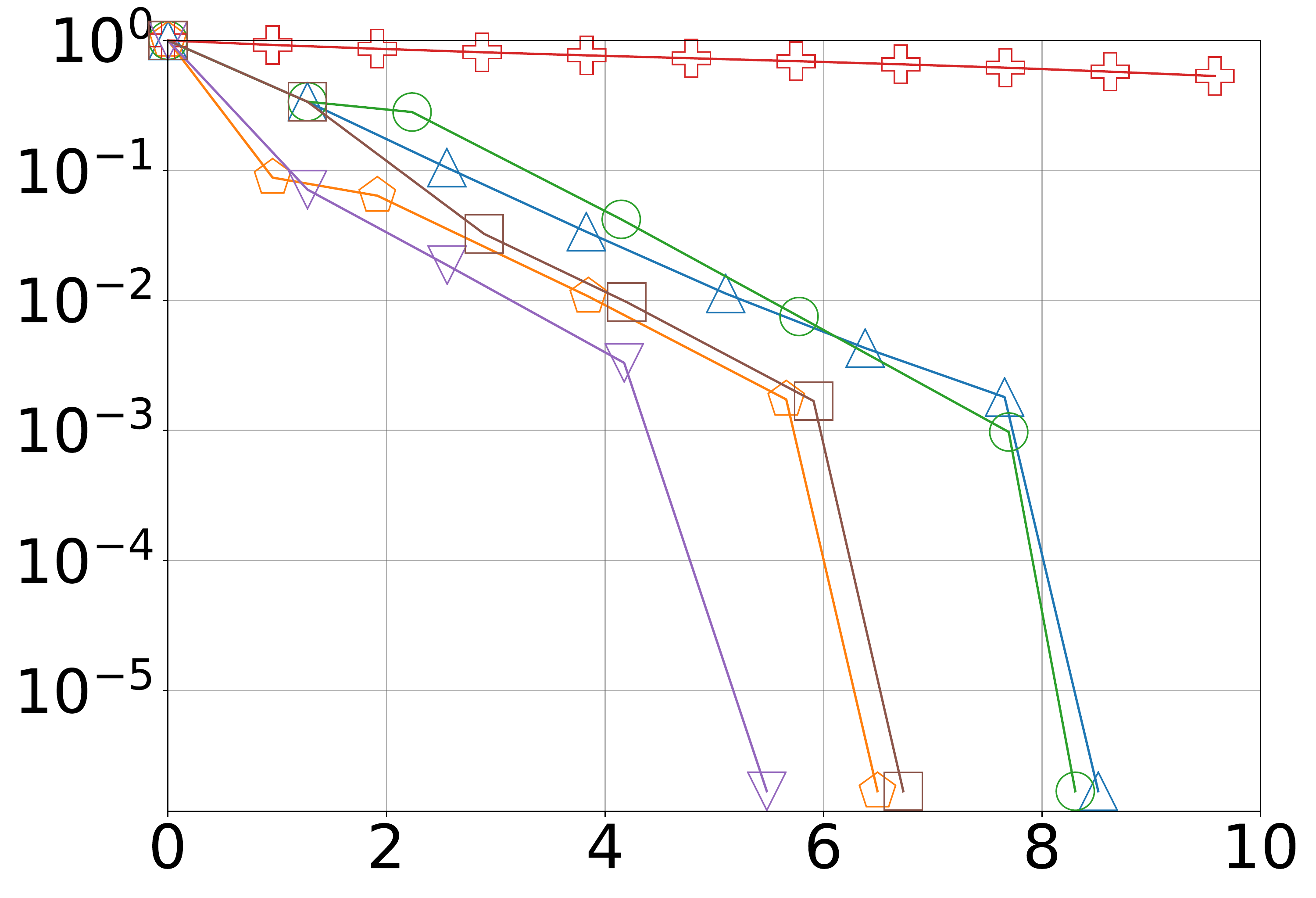} \\[-2mm]
	(a) {\em Dartmouth  1st Oct} & 
	(b) {\em Dartmouth  1st Nov} & 
	(c) {\em Dartmouth  1st Dec}  \\[1mm]
	\end{tabular}
    }
	\vspace{-3mm}
     \figcapup 
     \caption{The Untracked Item Percentage v.s. Communication Cost ($10^3$x) on Dartmouth datasets} 
     \label{fig:untracked_percent_vs_communication_real_Dartmouth}
     \figcapdown
     \vspace{-1.5mm}
\end{figure*}

\begin{figure*}[ht!]
	 \vspace{-2mm}
 \centering
    \resizebox{1.0\textwidth}{!}{%
	 \begin{tabular}{ccc} 
	\includegraphics[width=0.26\linewidth]{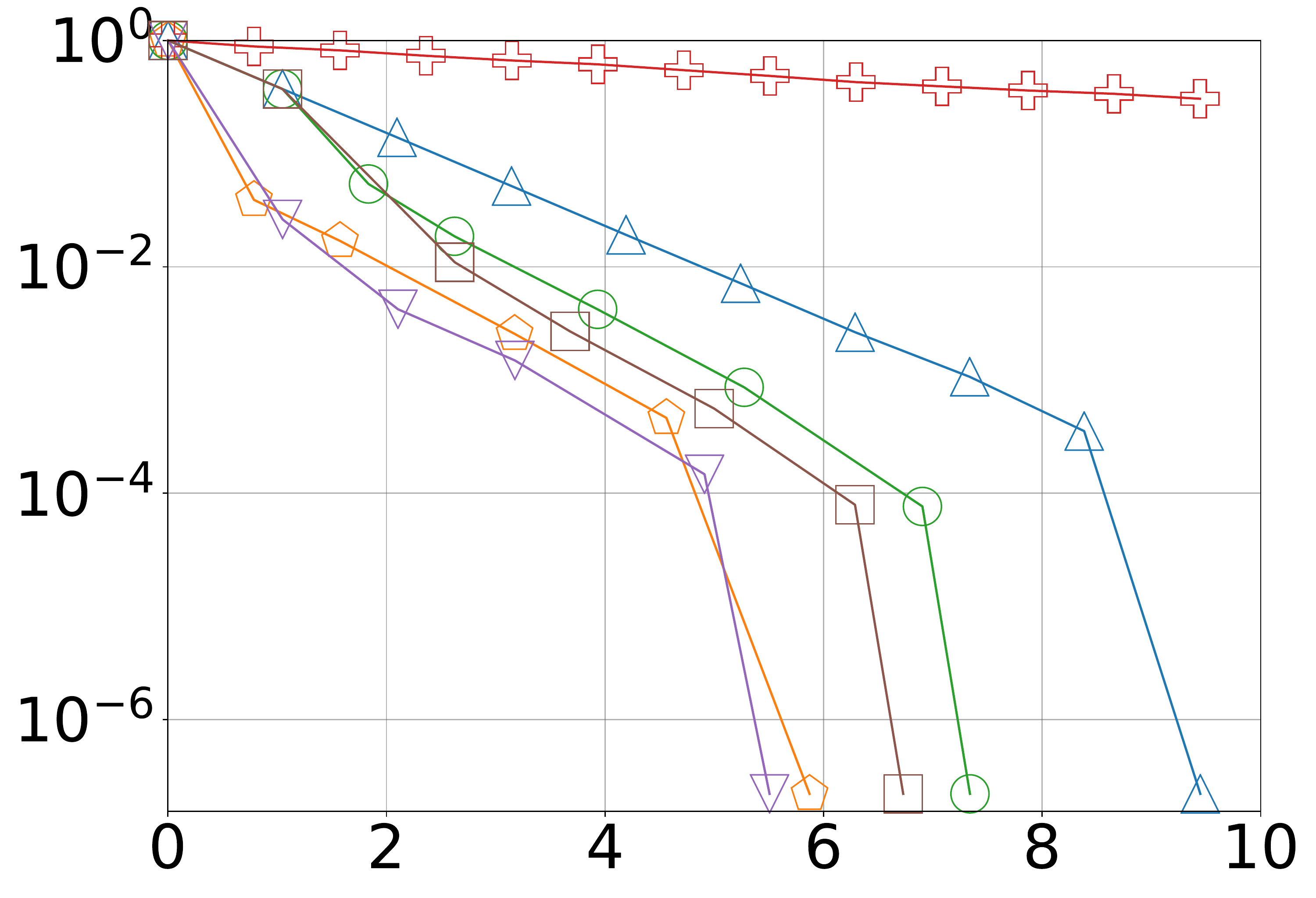} & 
	\includegraphics[width=0.26\linewidth]{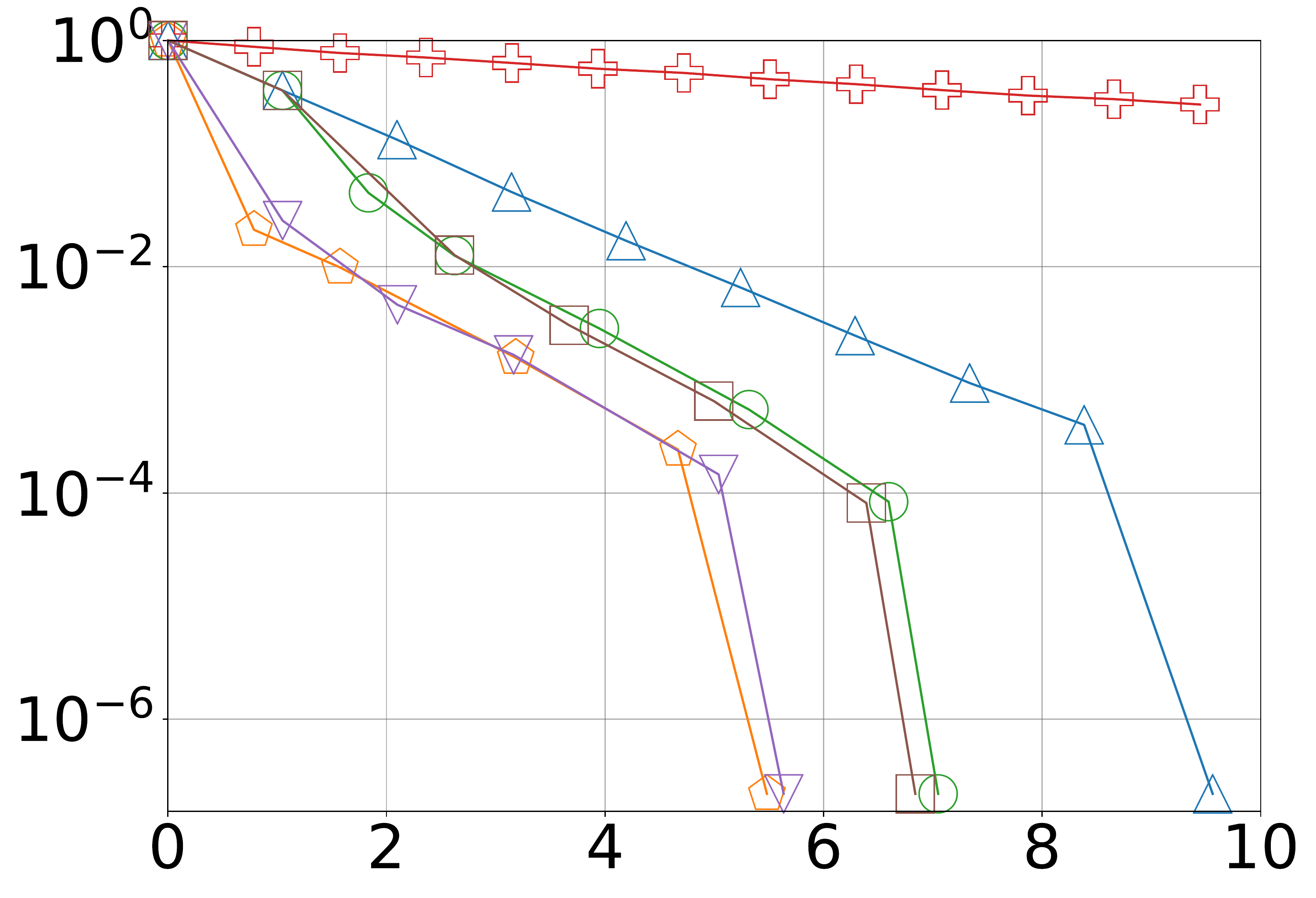} & 
	\includegraphics[width=0.26\linewidth]{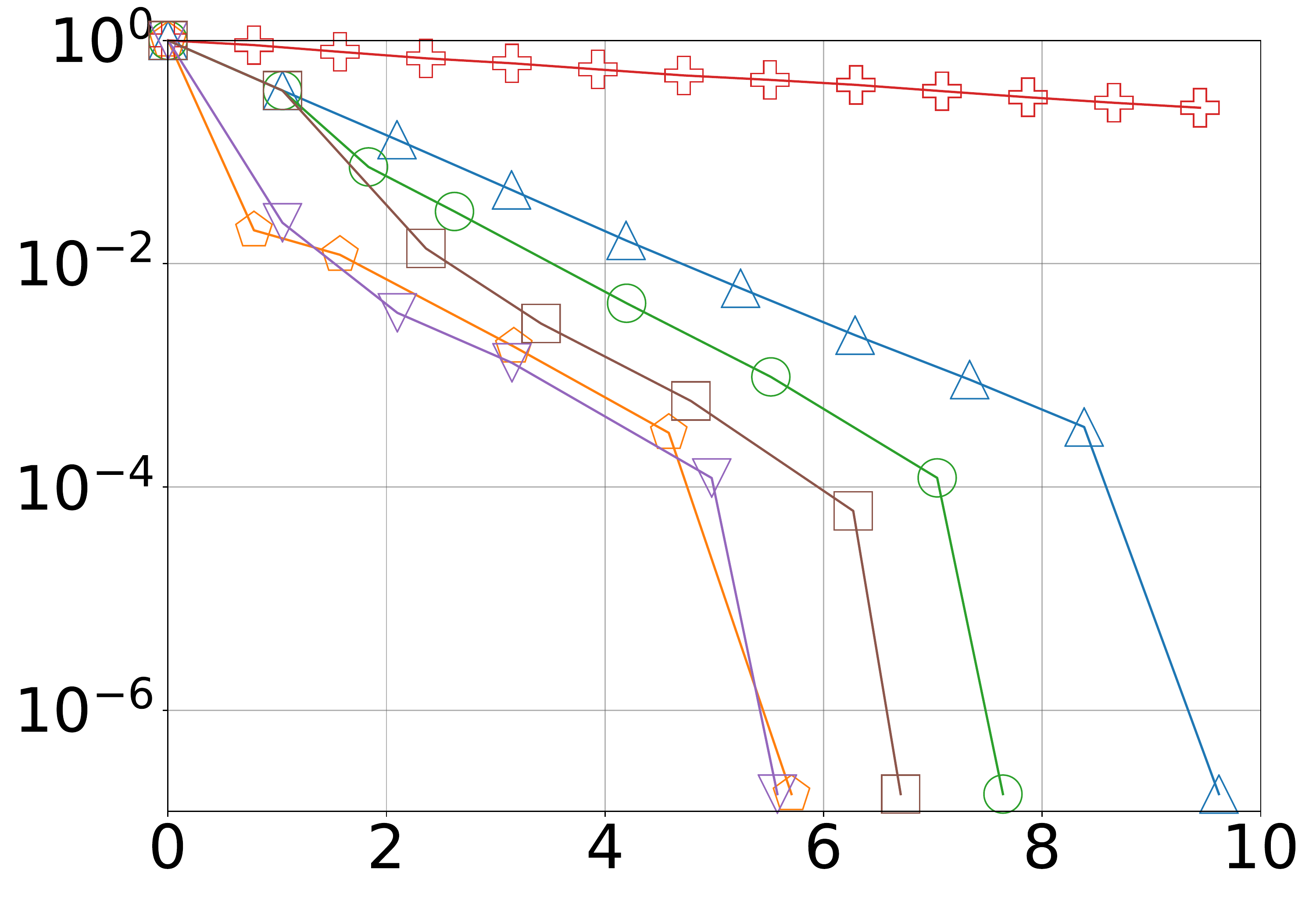} \\[-2mm]
	(a) {\em Uber Feb} & 
	(b) {\em Uber Apr} & 
	(c) {\em Uber June}  \\[1mm]
	
	\end{tabular}
    }
	\vspace{-3mm}
     \figcapup 
     \caption{The Untracked Item Percentage v.s. Communication Cost ($10^3$x) on Uber datasets} 
     \label{fig:untracked_percent_vs_communication_real_Uber}
     \figcapdown
     \vspace{-2mm}
\end{figure*}

\vspace{1mm}
\noindent \textbf{Real Datasets.} Below are the three real datasets we used. In each of the real datasets, we assign a unique id to each player (randomly and uniquely) in $[0, k)$, where $k$ is the number of players in the corresponding dataset.

\vspace{1mm}
\noindent\textit{World Cup HTTP request data}\footnote{\url{ftp://ita.ee.lbl.gov/html/contrib/WorldCup.html}}. The dataset consists of 92 days' requests to the 1998 World Cup website servers between April 30, 1998 and July 26, 1998. 
We use the requests of three representative days, namely the 30-th, the 60-th and the 90-th day, as the datasets in our experiment. On each selected day, the players are the servers that have received at least one request and the threshold to track is the number of requests on that day. Each item is a request that arrives at some server, in ascending order according to its time stamp.

\vspace{1mm}
\noindent
\textit{Dartmouth Campus Snmp Traceset}\footnote{\url{https://crawdad.org/dartmouth/campus/20090909/snmp}}~\cite{dartmouth-campus-20090909}. The dataset contains polling records of access points (AP) at Dartmouth College by Simple Network Management Protocol (SNMP) in Fall 2001. We use the records in three days, i.e., 1st Oct, 1st Nov, and 1st Dec. Each AP is reviewed as a player and the number of polling records on the selected day as the threshold. Each polling record is an item arriving in ascending order by its time stamp.

\vspace{1mm}
\noindent
\textit{Uber Pickups}\footnote{\url{https://www.kaggle.com/fivethirtyeight/uber-pickups-in-new-york-city$\#$uber-raw-data-janjune-15.csv}}. This dataset contains data on over the Uber pickups in New York City from January to June, 2015. Each record has a pickup time, a pickup location id and some other information. We take the pickups in three months, i.e., February, April and June as datasets. We consider the locations as players and the threshold to report is the number of pickups within the corresponding month. Each pickup record is treated as an item arriving in ascending order by the pickup time. 

\noindent \textbf{Synthetic Datasets.} The data in the synthetic datasets are generated with various distributions. Specifically, the distributions are: Uniform, Gaussian, Zipfian and Exponential. The number $k$ of the players varies from $2$ to $256$ (with multiplicative factor 2) and the threshold $N$ ranges from $2^{10} $ to $2^{24}$. 
Moreover, for each generated value $x$, if $x$ is not $[0, k)$, then we just simply discard it; on the other hand, if $x$ is not an integer,
we take its floor, i.e., $\lf x \rf$ to round it into a player id. 
In particular, the parameters of each of distribution are as follows:
\begin{itemize}[leftmargin = *]
	\item {\bf Uniform}: the id of the player for each item is generated uniformly at random in $[0, k)$. 
	\item {\bf Gaussian}: we set the mean to $k/ 2$ and standard variance to $k/6$;
	\item {\bf Zipfian}: each item arrives at player $i$ with probability \emph{proportional}\footnote{By proportional we mean here: the probability is normalized subject to the condition that the sum of the probabilities corresponding to the players is 1.} to $\frac{1}{\sqrt{i+1}}$;
	\item {\bf Exponential}: each item has probability proportional to $\exp(-i)$ arriving at the $i$-th player.
\end{itemize}

\vspace{1mm}
\noindent \textbf{Parameter Settings.} 
As the combination of all the parameters and the distributions is considerably large, we set the default values of $N$ to $2^{20}$ and of $k$ to $16$. When varying a parameter, the other is set to its default value. 
Furthermore, the $\DstT$ and $\DstDT$ require to know the concrete distribution of the datasets as an input, which may be unavailable for real datasets. Thus, we use \emph{frequencies} of each players as its multinormial distribution. 
Finally, we set the failure probability to $1\%$, which suffices for most of the applications in practice.

\subsection{Results on the Real Datasets}

Figure~\ref{fig:untracked_percent_vs_communication_real_WorldCup}-\ref{fig:untracked_percent_vs_communication_real_Uber} illustrate the results of the algorithms on three real datasets. All plots in the figures refer to the percentage of untracked items as a function of the number of used communications. Each plot represents a round. Different algorithms require different number of communications for a round. Some plots are truncated for $\US$ because it takes much more communications that others. 

The figures show several results. First all algorithms, {$\DstT$} and {$\DstDT$} perform the best, with { $\DstDT$} slightly better in most cases. This is as expected because they know the frequency information and have more knowledge than the other algorithms. It confirms the effectiveness of our strategy. Compared to { $\DT$}, the number of communications reduces by $25\%$ (Figure~\ref{fig:untracked_percent_vs_communication_real_Dartmouth} (b)) to $75\%$ (Figure~\ref{fig:untracked_percent_vs_communication_real_WorldCup} (c)). 

Second, the performances of {$\LnDstT$} and { $\LnDstDT$ } are inferior to { $\DstT$/$\DstDT$} but better than {$\DT$} in general. Compared to the { $\DstT$/$\DstT$}, they don't know the item arrival frequency at each player in the datasets and therefore have less information. They run {$\DT$} for the first round to learn an approximate distribution of the dataset. Therefore, their performance in the first round is exactly the same as {$\DT$} in the first round. The learned distribution helps in tracking the incoming items in most cases. Compared to { $\DT$}, we observe much sharper decreases in the curves from the second round. 

Third, both { $\DstDT$} and { $\LnDstDT$} exhibit better performance than { $\DstT$} and { $\LnDstT$}. As the real datasets do not necessarily have perfect distribution, incorporating {$\DT$} to handle counters' deviation from their expectation values in an aggregate and dynamic manner is more stable than using merely predetermined and static slacks. The only exception is the dataset \textit{WorldCup Day 90} (Figure~\ref{fig:untracked_percent_vs_communication_real_WorldCup} (c)), in which there are only two players and the distribution is rather skew and stable. Therefore, {$\DstT$} and { $\LnDstT$} win with static slacks.

The figures also show the effectiveness of our backup mechanism (as described in Appendix~\ref{appendix:opt}) when the distribution of real dataset is not stable. In \textit{Dartmouth} datasets, the distribution is rather unstable -- {$\DstT$} and {$\LnDstT$} fail in the second round and track much fewer percentage of items in the second round than the first one. Detecting the degeneracy in efficiency, they switch to $\DT$ in the third round. 
{ $\LnDstT$} lose only by a marginal amount to {$\DT$} even in this case.

Finally, the {$\US$} algorithm gives the worst performance as its time complexity grows quadratically with respect to $k$, the number of players. Further, it is sensitive to skew distortions. In \textit{WorldCup Day 60} (Figure~\ref{fig:untracked_percent_vs_communication_real_WorldCup} (b)), {$\US$} exhibits frequent termination of rounds as the tailing items comes in a very unbalanced manner. All other algorithms have switched to { $\DT$} and handle the tailing items smoothly. 

\subsection{Results on the Synthetic Datasets}

\begin{figure*}[t!]
	 \vspace{-2mm}
	 \includegraphics[width=1.0\linewidth]{IMG/legend} \\
	\begin{tabular}{cccc} 
    \hspace{-6.5mm} 
	\includegraphics[width=0.26\linewidth]{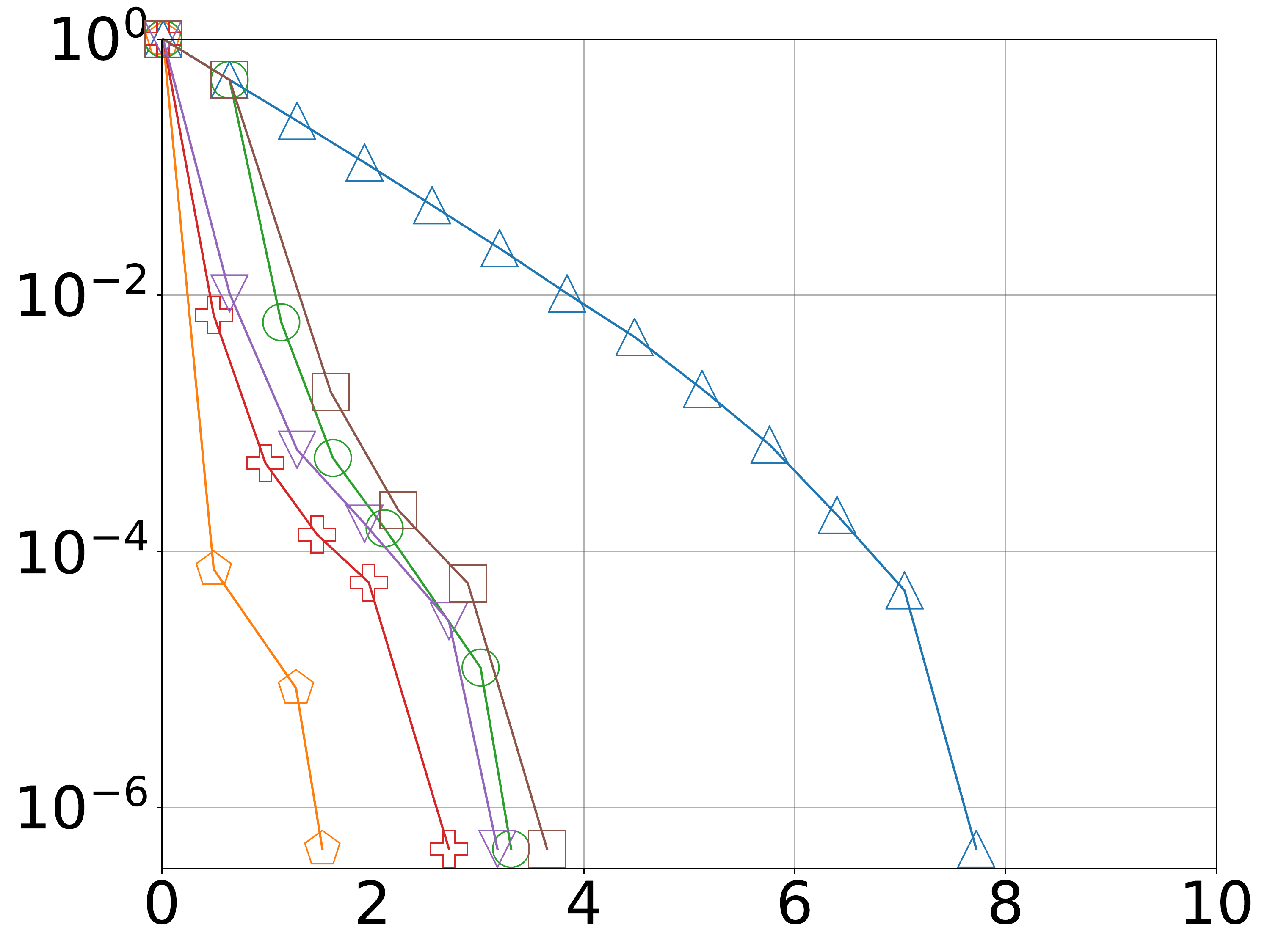} &     
	\hspace{-5mm} 
	\includegraphics[width=0.26\linewidth]{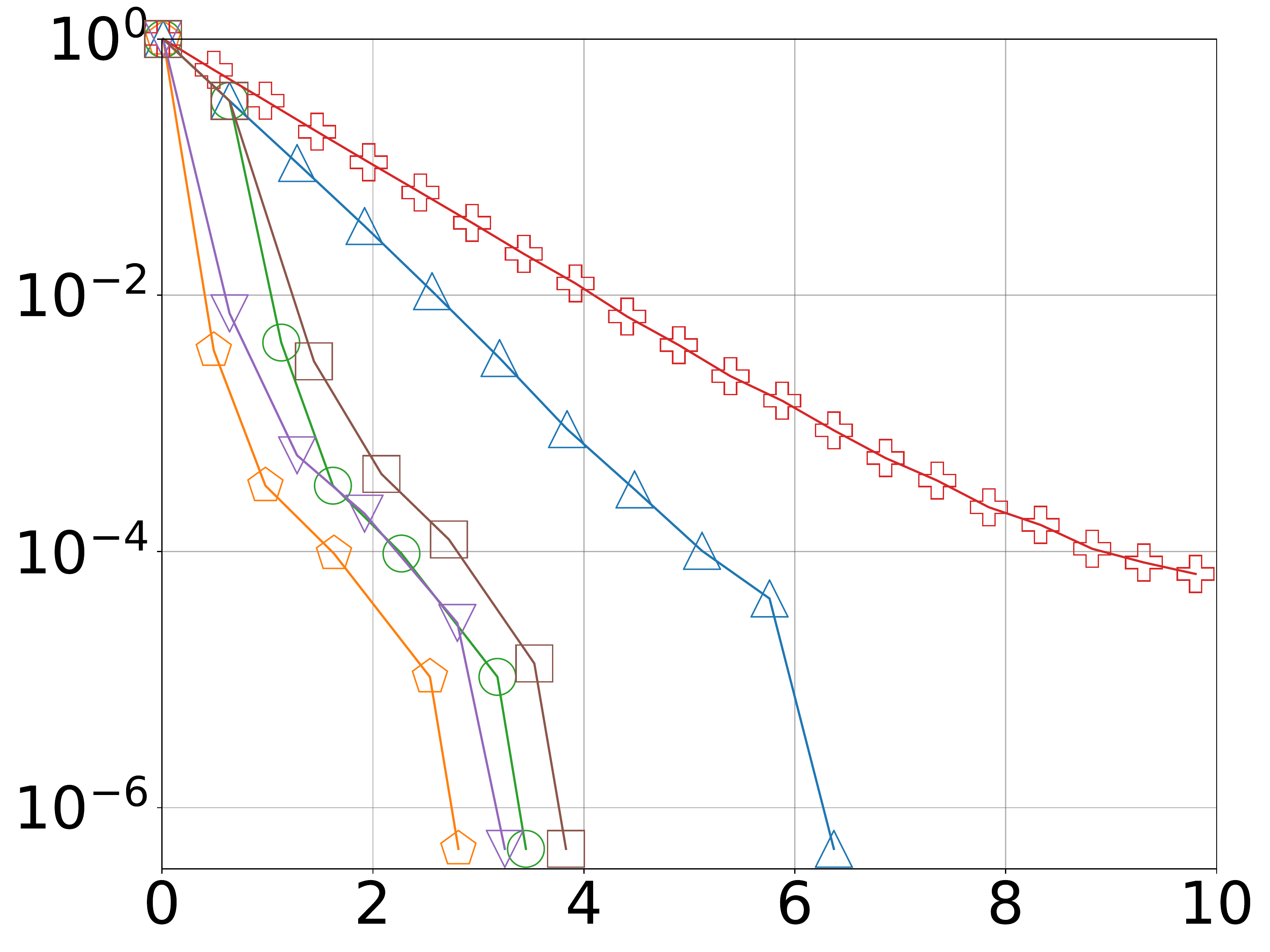} & 
	\hspace{-5mm} 	
	\includegraphics[width=0.26\linewidth]{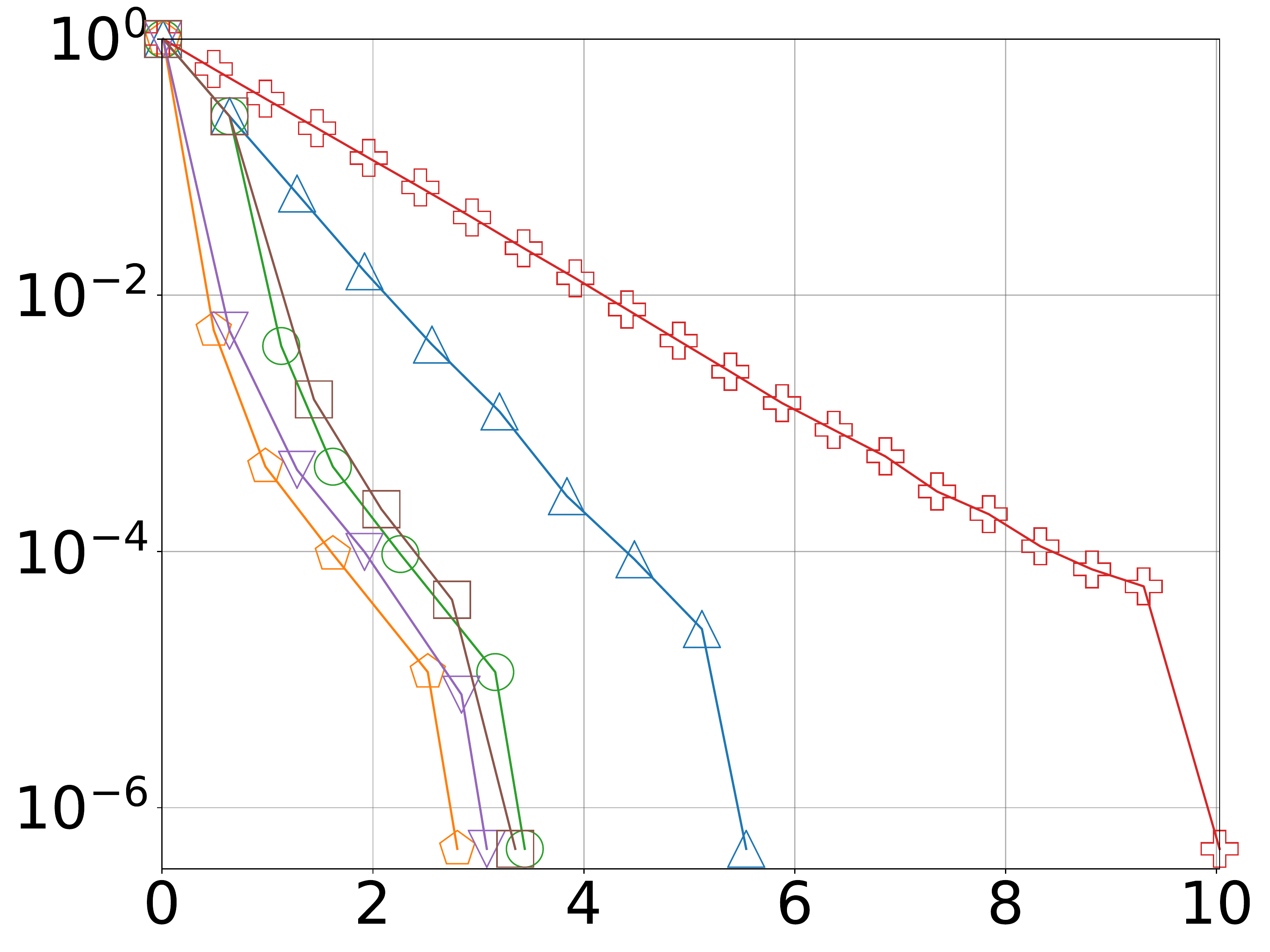} & 	
	\hspace{-5mm} 
	\includegraphics[width=0.26\linewidth]{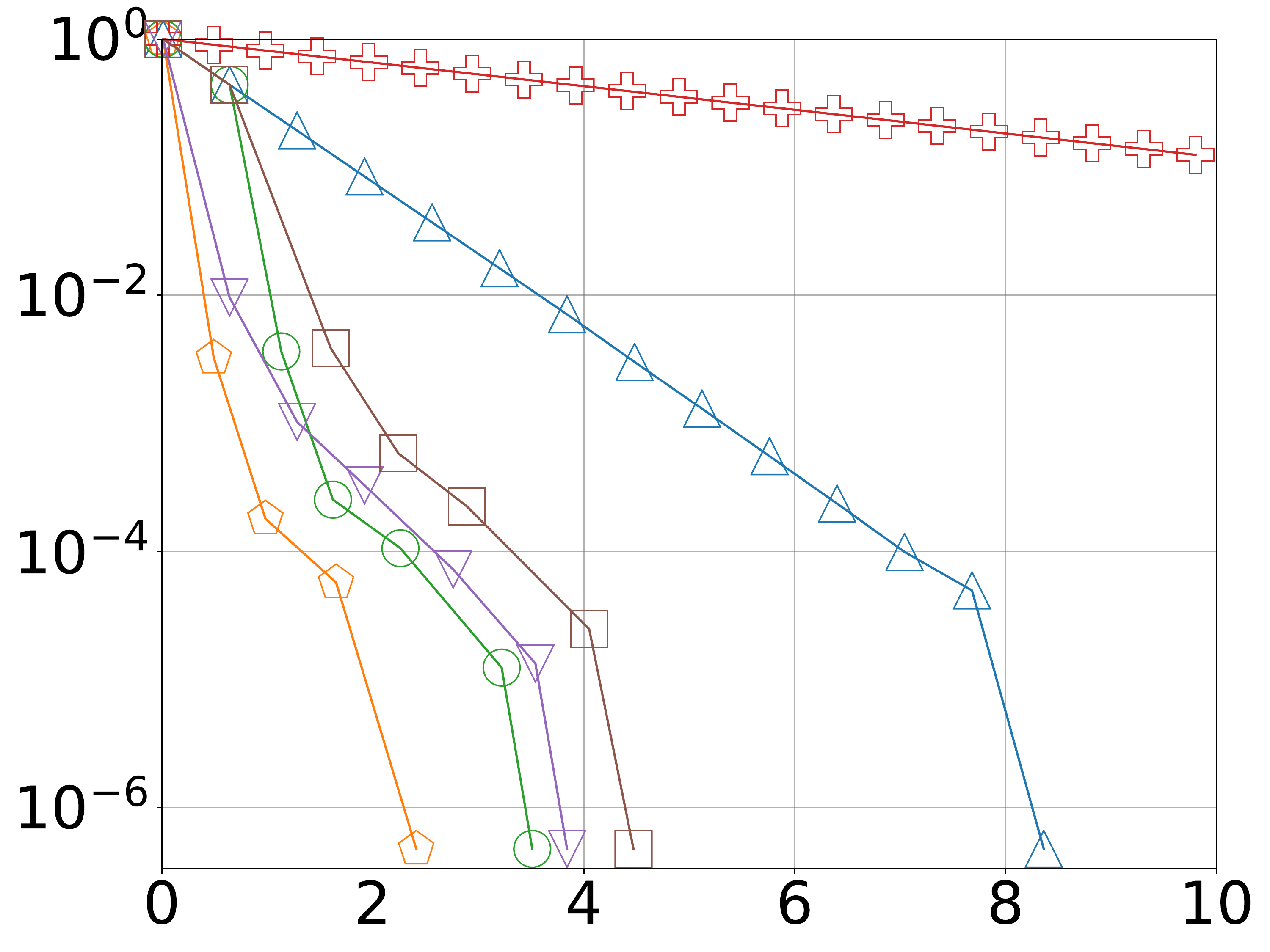} \\
	\hspace{-10mm} (a) {\em Uniform} & \hspace{-4.6mm} (b) {\em Gaussian} & \hspace{-4.6mm} (c) {\em Zipfian}  & \hspace{-4.6mm} (d)  {\em Exponential}\\[1mm]
	\end{tabular}
	\vspace{-3mm}
     \figcapup 
     \caption{The Untracked Item Percentage v.s. Communication Cost ($100$x) on synthetic datasets} 
     \label{fig:untracked_percent_vs_communication_synthetic}
     \figcapdown
 \end{figure*}

\begin{figure*}[t!]
	 \vspace{-2mm}
	 \begin{tabular}{cccc} 
    \hspace{-5mm} 
    \includegraphics[width=0.25\linewidth]{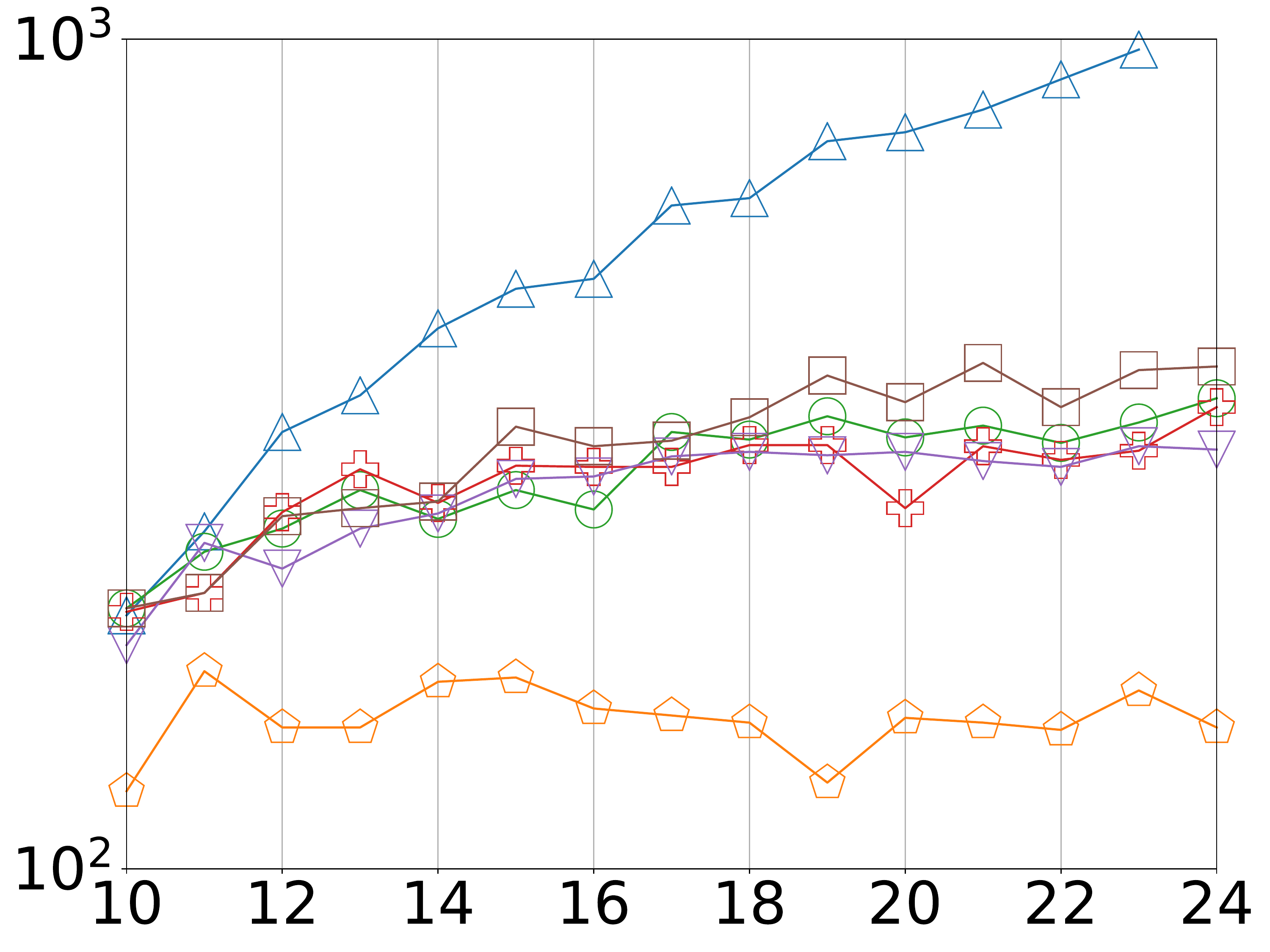} & 
   \hspace{-3mm}
	\includegraphics[width=0.25\linewidth]{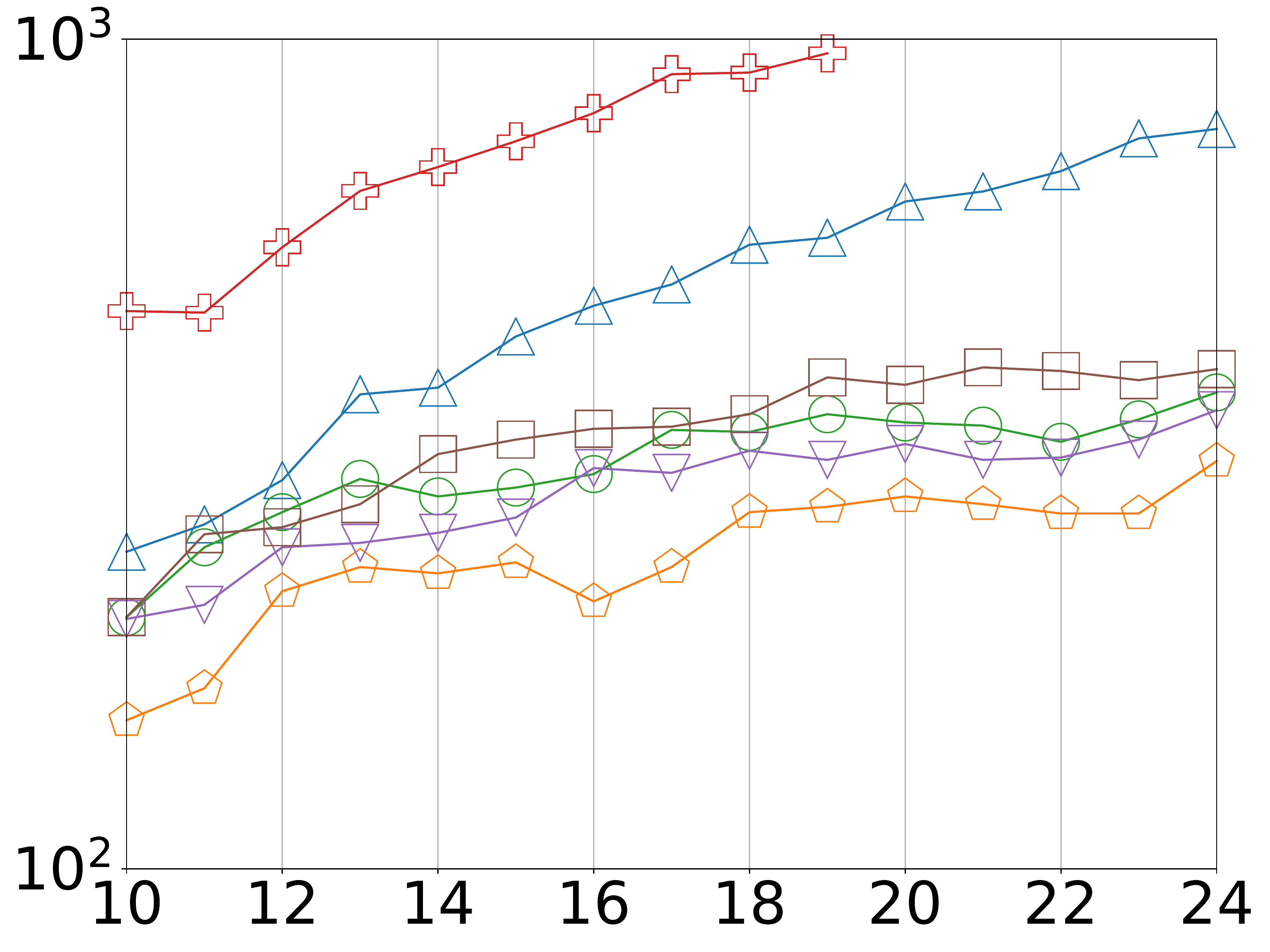} & 
    \hspace{-3mm}
	\includegraphics[width=0.25\linewidth]{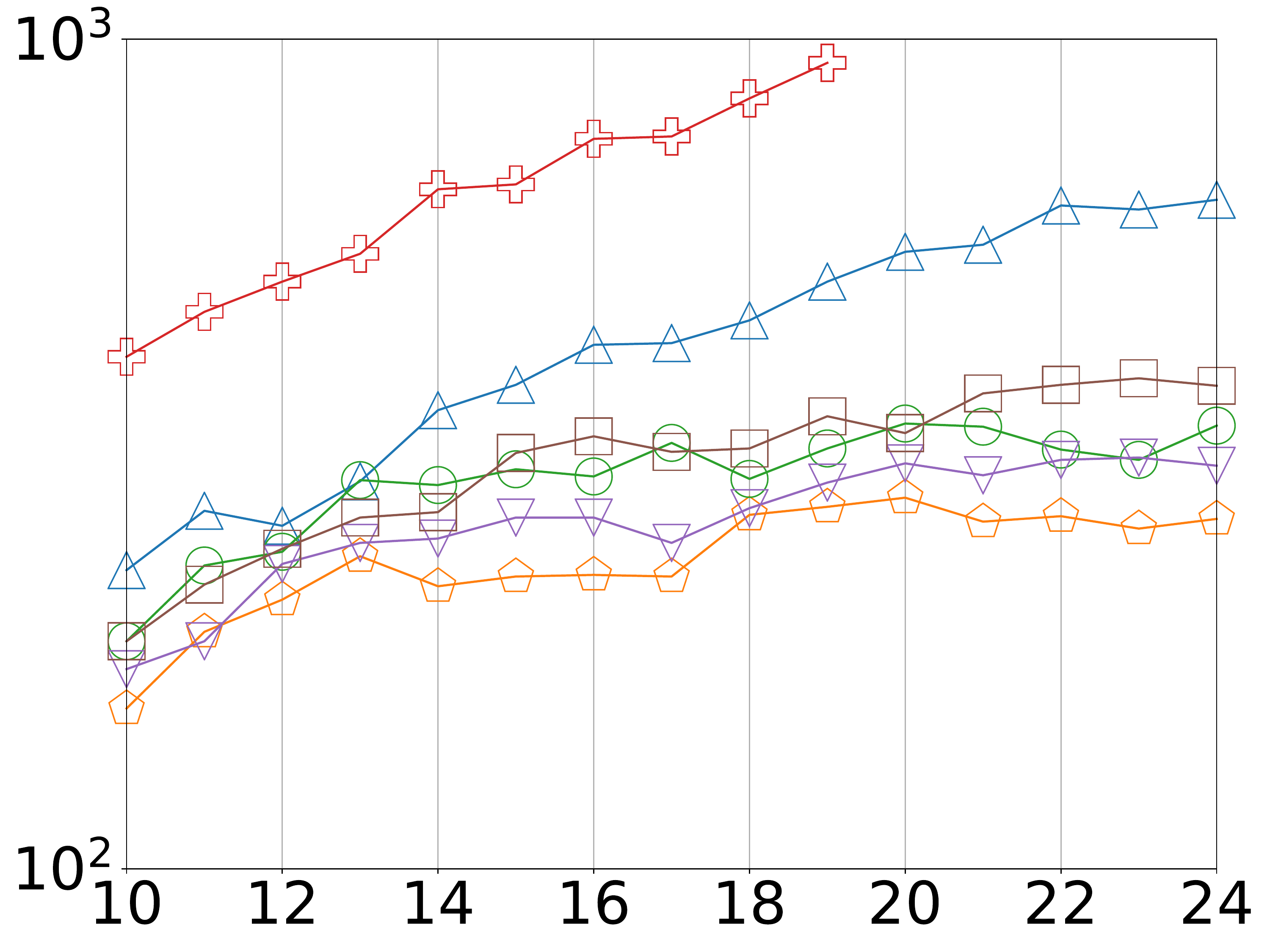} & 
    \hspace{-3mm}
	\includegraphics[width=0.25\linewidth]{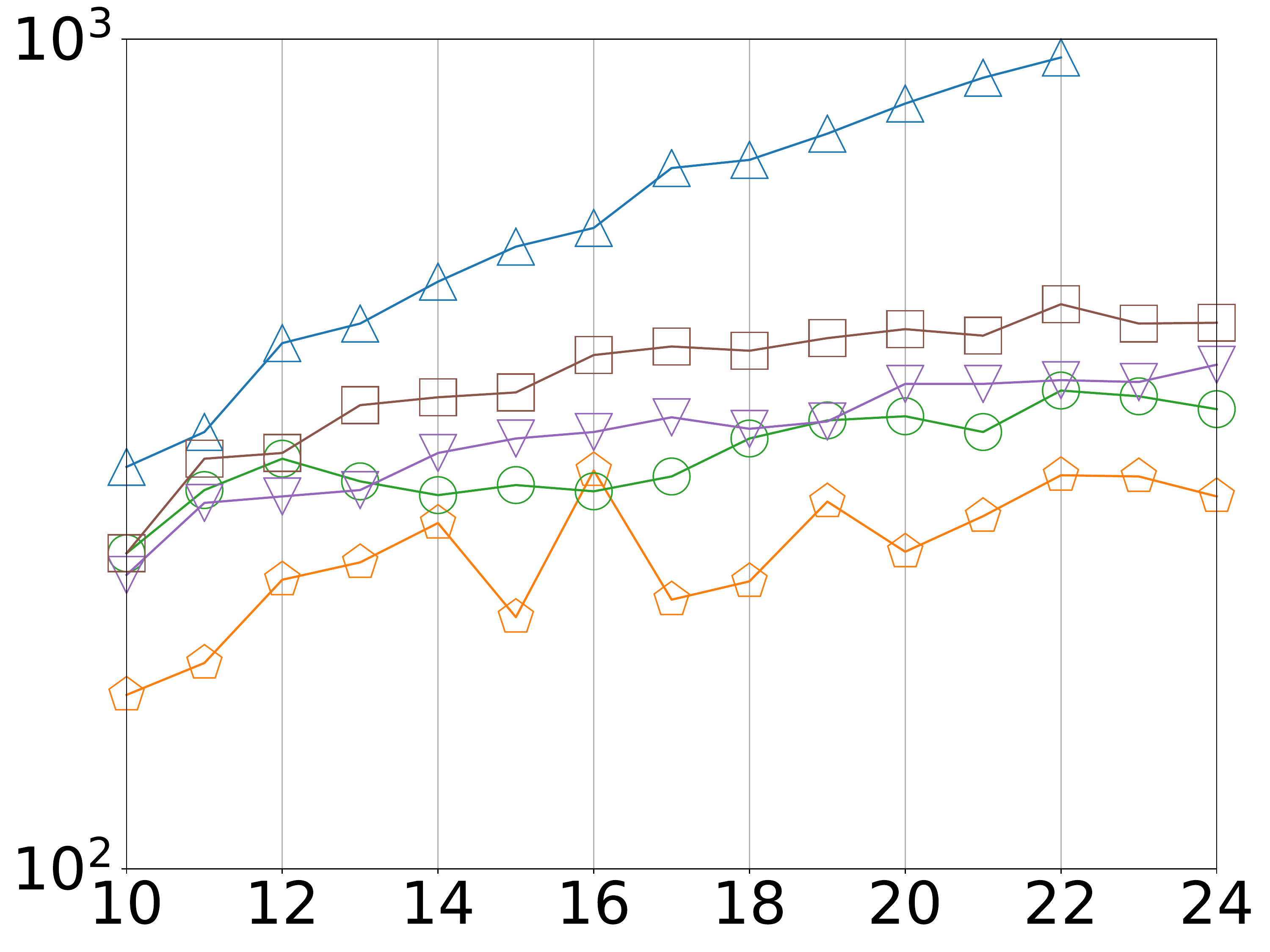} \\
	\hspace{-10mm} (a) {\em Uniform} & \hspace{-4.6mm} (b) {\em Gaussian} & \hspace{-4.6mm} (c) {\em Zipfian}  & \hspace{-4.6mm} (d)  {\em Exponential}\\[1mm]
	
	\end{tabular}
	\vspace{-3mm}
     \figcapup 
     \caption{Communication Cost v.s. $N$ ($2^x$) on synthetic datasets} 
     \label{fig:communication_vs_threshold}
     \figcapdown
	\vspace{-3mm}
 \end{figure*}

 \begin{figure*}
	 \vspace{-2mm}
	 \begin{tabular}{cccc} 

	\hspace{-6.2mm} 
	\includegraphics[width=0.27\linewidth]{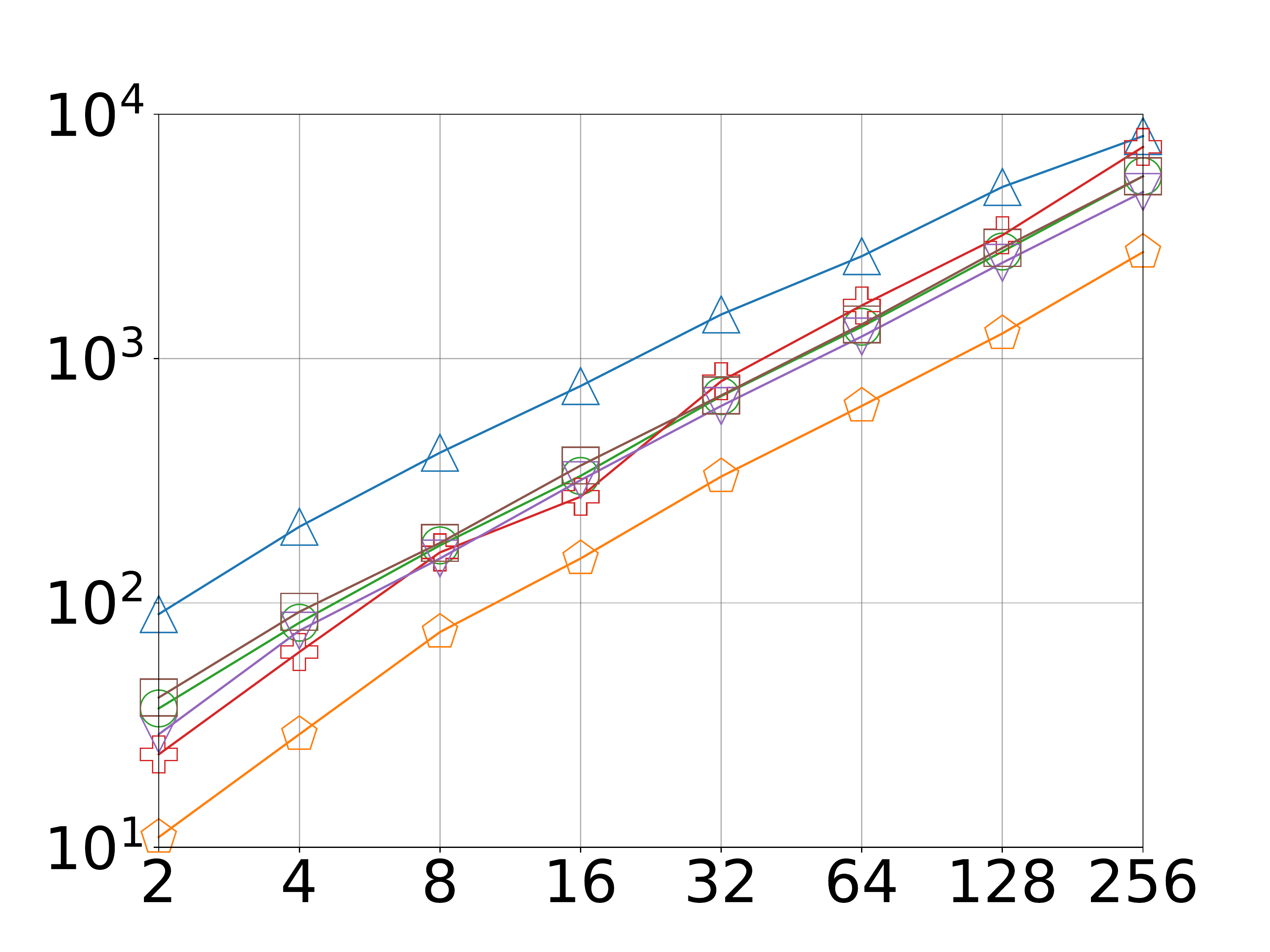} & 
   \hspace{-6.5mm}
    \includegraphics[width=0.27\linewidth]{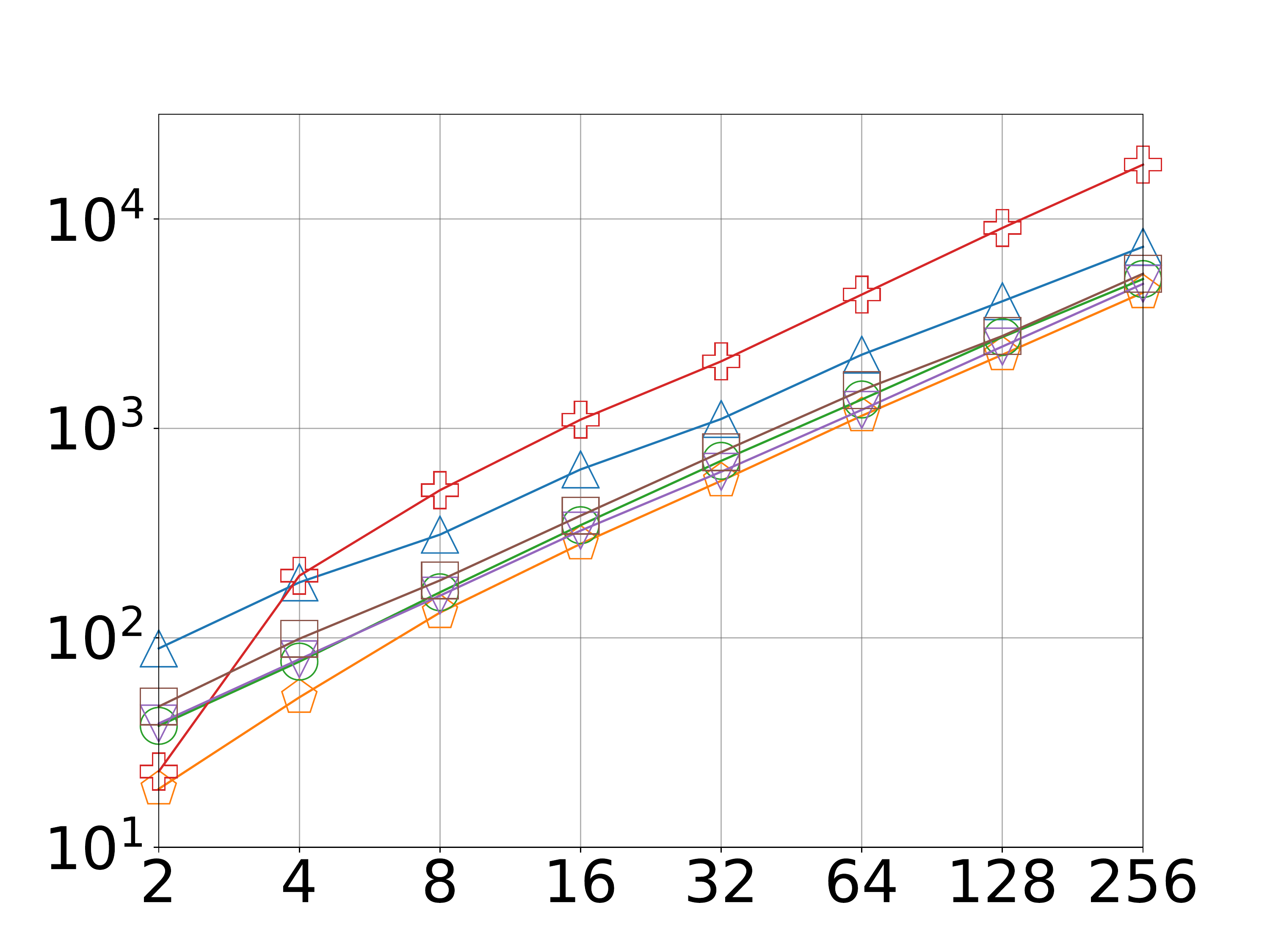} & 
   \hspace{-6mm}
	\includegraphics[width=0.27\linewidth]{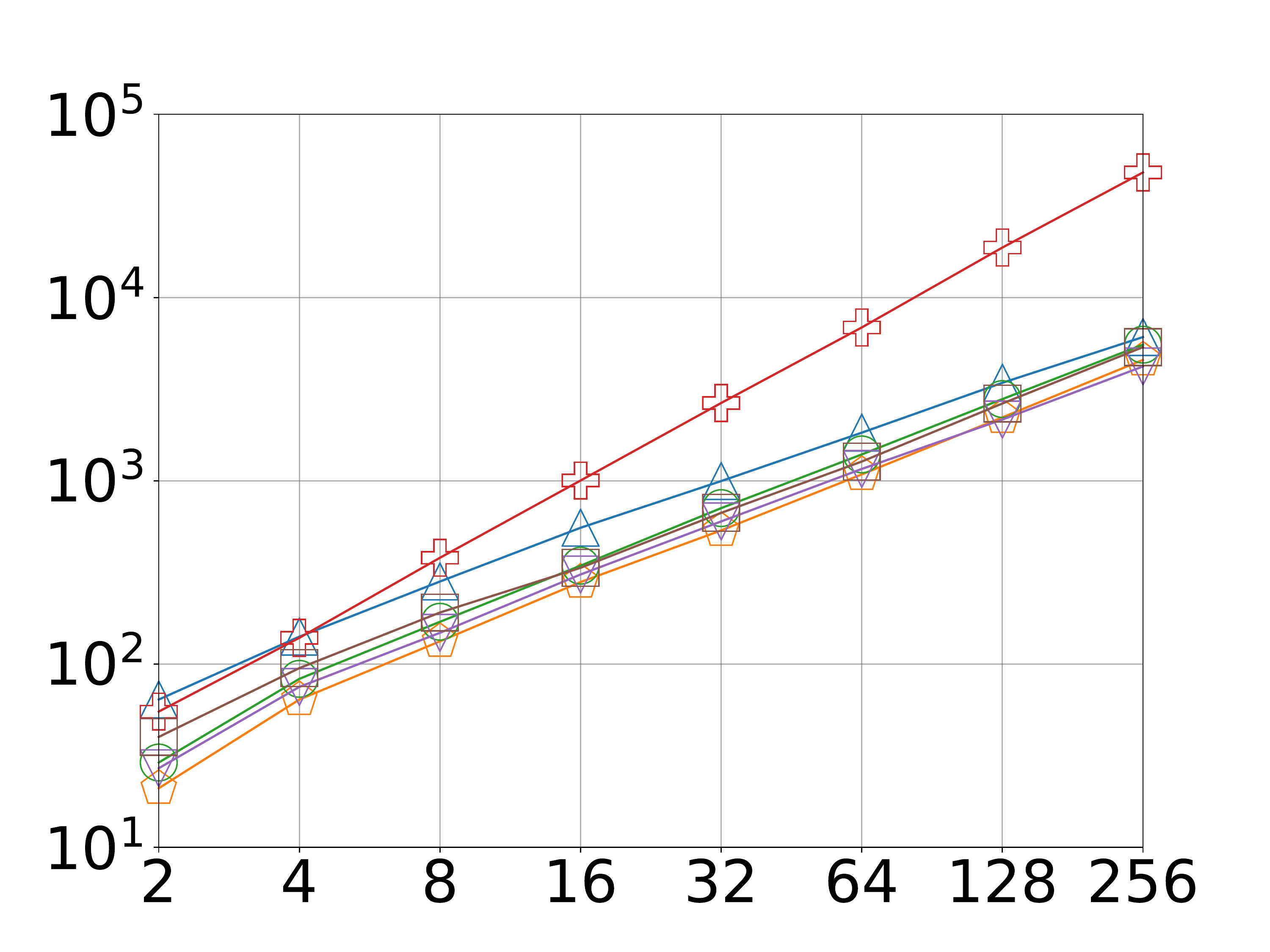} & 
   \hspace{-5mm}
	\includegraphics[width=0.27\linewidth]{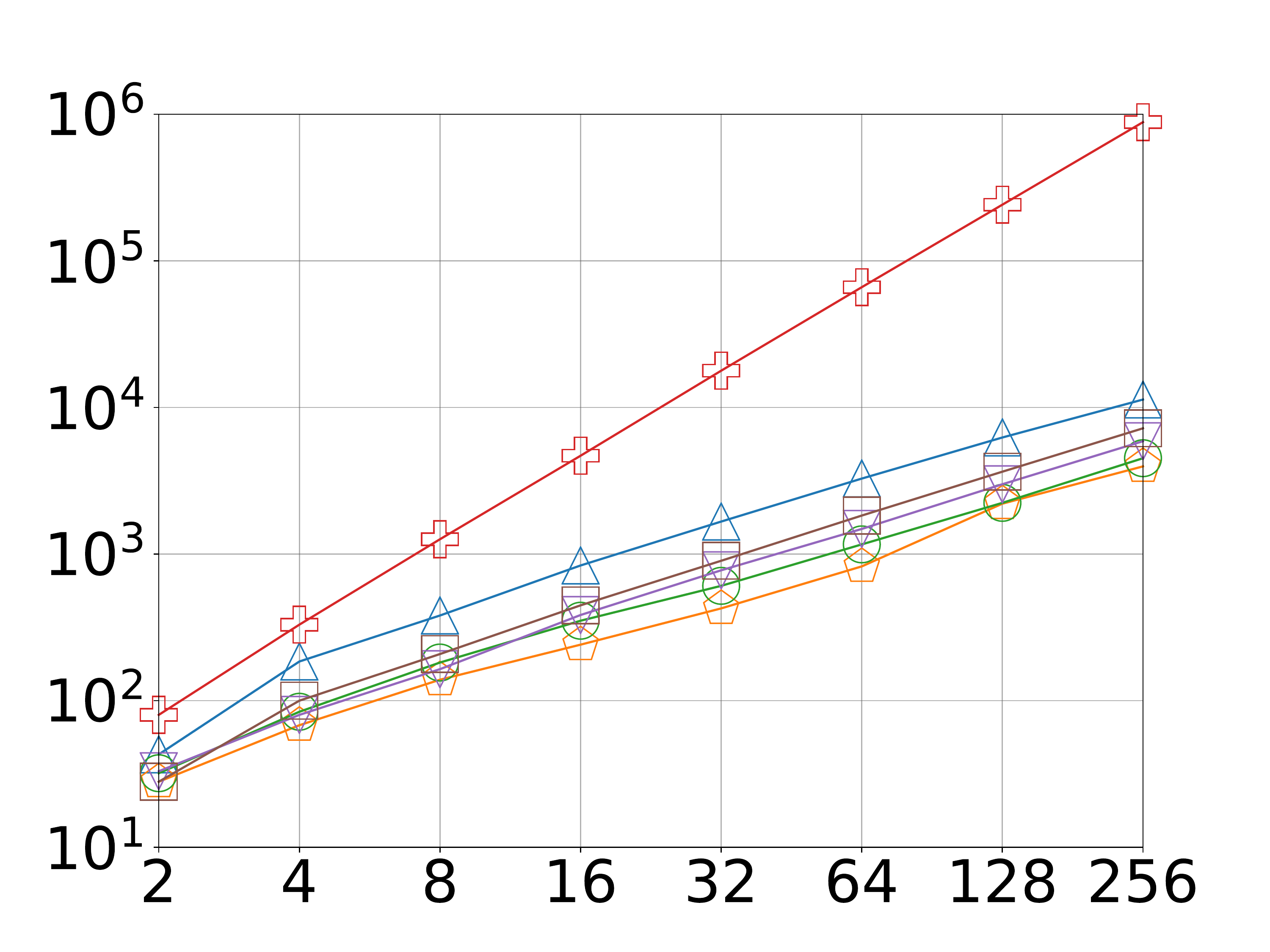} \\
	\hspace{-10mm} (a) {\em Uniform} & \hspace{-4.6mm} (b) {\em Gaussian} & \hspace{-4.6mm} (c) {\em Zipfian}  & \hspace{-4.6mm} (d)  {\em Exponential}\\[1mm]
	
	\end{tabular}
	\vspace{-3mm}
     \figcapup 
     \caption{Communication Cost v.s. $k$ on synthetic datasets} 
     \label{fig:communication_vs_player}
	\vspace{-3mm}
     \figcapdown
 \end{figure*}

\noindent \textbf{Sensitivity to Distribution.} Figure~\ref{fig:untracked_percent_vs_communication_synthetic} shows the efficiency of the algorithms under various distributions. It plots the percentage of untracked items as a function of the number of communications. Each plot represents one round. When the datasets are generated from some distribution, our algorithms perform consistently better than the { $\DT$} algorithms, regardless of the distribution. The communication is reduced by $33\%$ to $80\%$. Furthermore, our algorithms with static slacks perform better than their {$\DT$} counterparts. When the distribution is stable, the static slacks capture more accurately the number of items a player will receive. 

\vspace{2mm}
\noindent \textbf{Sensitivity to Threshold.} 
Figure~\ref{fig:communication_vs_threshold} plots the number of communications as a function of the $N$ (the threshold) under the various distributions, with the number of players fixed to default value $16$. We truncate the plots with cost more than $10^3$. {$\US$} performs really good on Uniform distribution as each player receives roughly the same number of items. But its communication cost blows up on other datasets. It does not show up in the plot for \textit{Exponential} distribution because it uses more than $10^3$ communication even for $N = 1k$. Moreover, the figures shows an increasing efficiency gain of our algorithms compared to {$\DT$}, as $N$ increases. This is consistent with our analysis of their communication complexity. While $\DstT$ offers the best performance over all the datasets, $\DstDT$, $\LnDstT$ and $\LnDstDT$ yield comparable performance. 

\vspace{2mm}
\noindent \textbf{Sensitivity to Number of Players.} Figure~\ref{fig:communication_vs_player} plots the number of communications as a function of the number of players under the various distributions, with the threshold fixed to default value $1m$. We truncate the result for {$\US$} when its communication exceeds $10^4$. The figures illustrate a shrinking gap in the number of communications between $\DT$ and the our algorithms as the number of players increases. This complies with our theoretical analysis as the ratio of the communication complexity between the two is given by $O((k \log \frac{N}{k}) / (k \log \log \frac{N}{k \ln \frac{k}{\delta}} + k \log \log \frac{k}{\delta})) = O((\log \frac{N}{k}) / (\log \log \frac{N}{k \ln \frac{k}{\delta}} + \log \log \frac{k}{\delta}))$. When $N$ and $\delta$ are fixed, the ratio decreases as $k$ increases. The only exception is the dataset with Exponential distribution and with 2 players, in which the { $\DT$} perform very well. In such case the number of items the first player receives is roughly $\exp(1)$ times that of the second player. A round terminates after player one sending two notifications to the coordinator (the slack size is $s = n_1 / 2$). On the other hand, player two is assigned the same slack $s = n_1 / 2$ but receives $n_1 / \exp(1)$ items. Only a small fraction of the slack is wasted in this case. 

\section{Conclusion}

The paper exploits the counter increment distribution and presents four data-dependent algorithms that utilize knowledge on the data distribution. All our algorithms have communication cost $O(k \log \log \frac{N}{k \ln \frac{k}{\delta}} + k \log \log \frac{k}{\delta} )$, where $\delta$ is a parameter controls the failure probability, improving the state-of-the-art  $O(k \log \frac{N}{k})$ bound. 
In addition, our algorithms are equipped with backup mechanism that guarantees comparable performance as the data-independent $\cmy$ algorithm  when the distribution fluctuates. 
We experimentally evaluate our algorithms against the state-of-the-art competitors, using both real and synthetic datasets. 
Our experimental results show the efficiency and robustness of our algorithms. 

\begin{acks}
	Junhao Gan is supported by Australian Research Council (ARC) DECRA DE190101118.
\end{acks}

\bibliographystyle{ACM-Reference-Format}
\bibliography{ref}

\newpage
\appendix

\section{Proofs of Lemmas and Theorems} \label{APPENDIX_PROOFS}

\noindent \textbf{Proof of Lemma~\ref{lma:setting UB}}.
Denote $X = n_i = \sum_{j \in [t]} X_j$. As $E[X_j] = \mu_i$, by linearity of expectation and independence, we have $E[X] = t \mu_i$. Further, as the $X_j$'s are independent and $Var[X_j]=\sigma^2 = \mu_i (1 - \mu_i)$, $Var[X] = t \sigma^2$. Applying Fact~\ref{lma:Bernstein inequality} with $Y_j = X_j$, $M = 1$ and setting the failure probability to $\frac{\delta}{k}$, we have 
\begin{equation*}
    \Pr[X - t\mu_i \ge \lambda] \le \exp{\left( - \frac{\lambda^2}{2(t \sigma^2 + \lambda / 3) } \right) } = \frac{\delta}{k}
\end{equation*}
It follows that 
\begin{equation*}
    \lambda^2 - \left( \frac{2}{3} \ln \frac{k}{\delta} \right) \lambda - 2t\sigma^2 \ln \frac{k}{\delta} = 0
\end{equation*}
Solving the quadratic equation gives
\begin{align*}
\lambda 
    &= \frac{1}{2} \left( \frac{2}{3} \ln \frac{k}{\delta} + \sqrt{\frac{4}{9} \ln^2 \frac{k}{\delta} + 8 t \sigma^2  \ln \frac{k}{\delta} }  \right) \\
    &\le \frac{2}{3} \ln \frac{k}{\delta} + \sqrt{2 t \sigma^2  \ln \frac{k}{\delta} }
\end{align*}
Therefore $UB_i \doteq t \mu_i + \frac{2}{3} \ln \frac{k}{\delta} + \sqrt{2 t \sigma^2  \ln \frac{k}{\delta} } \ge t\mu_i + \lambda$, and we have
\begin{equation*}
    \Pr[X \ge UB_i] \le \Pr[X \ge t\mu_i + \lambda] \le \frac{\delta}{k}
\end{equation*}
Similarly, if we take $Z_j = -X_j$, then $E[Z_j] = -\mu_i$ and $Var[Z_j] = Var[-X_j] = \sigma^2$. Let $Z = \sum_{j \in [t]} Z_j$, then we have 
\begin{equation*}
    \Pr \big[ Z \ge -t \mu_i  + \sqrt{2 t \sigma^2 \ln \frac{k}{\delta} } + \frac{2}{3} \ln \frac{k}{\delta} \big] \le \frac{\delta}{k}
\end{equation*}
which is equivalent to 
\begin{equation*}
      \Pr \big[ X \le t \mu_i  - \sqrt{2 t \sigma^2 \ln \frac{k}{\delta} } - \frac{2}{3} \ln \frac{k}{\delta} \big] \le \frac{\delta}{k}
\end{equation*}
This finishes the proof. 
\myqed

\vspace{2mm}
\noindent \textbf{Proof of Lemma~\ref{lma:setting t}}. 
Recall that $s_i = \mu_i t + \sqrt{2 t \mu_i (1 - \mu_i) \ln \frac{k}{\delta} } + \frac{2}{3} \ln \frac{k}{\delta} $. Summing over all players $i \in [k]$, we get

\begin{align*}
f(t) &= \sum_{i = 1}^k \mu_i t + \sum_{i = 1}^k \sqrt{2 t \mu_i (1 - \mu_i) \ln \frac{k}{\delta} } + \sum_{i = 1}^k  \frac{2}{3} \ln \frac{k}{\delta}  
\end{align*}
The first and third terms sum up to $t$ and $\frac{2k}{3} \ln \frac{k}{\delta}$ respectively. It is left to bound the second term. We utilize the concavity of the square root function $\sqrt{\cdot }$ and obtain
\begin{align} 
    \label{ineq:concave_of_root_1}
    \sum_{i = 1}^k \frac{1}{k} \sqrt{2 t \mu_i (1 - \mu_i) \ln \frac{k}{\delta} } 
    &\le \sqrt{\sum_{i = 1}^k \frac{1}{k}  \left( 2 t \mu_i (1 - \mu_i) \ln \frac{k}{\delta} \right) }  \\
    \label{ineq:concave_of_root_3}
    &\le \sqrt{ \frac{2 t}{k} \ln \frac{k}{\delta} } 
\end{align}    
The second inequality follows from $(1 - \mu_i) \le 1$ for $\forall i \in [k]$ and $\sum_{i = 1}^k \mu_i = 1$. Therefore, 
\begin{equation}
    \label{ineq:deriving_N_minus_sqrt_N_for_t_1}
    f(t) \le  t + \sqrt{ 2 k t \ln \frac{k}{\delta} }  + \frac{2k}{3} \ln \frac{k}{\delta} 
\end{equation}
Further, when the $\DstT$ does not run the $\cmy$ algorithm, it holds that $k \ln \frac{k}{\delta} \le N$. Hence $\frac{2k}{3} \ln \frac{k}{\delta}  \le \frac{2}{3} \sqrt{k N \ln \frac{k}{\delta} }$. Combining that $t \le N$, we have \begin{equation}
    \label{ineq:deriving_N_minus_sqrt_N_for_t_2}
    f(t) \le  t + \sqrt{ 2 k N \ln \frac{k}{\delta} }  + \frac{2}{3} \sqrt{k N \ln \frac{k}{\delta} }
\end{equation}
It suffices to take $t = N - (\sqrt 2 + 2/3)\sqrt{k N \ln \frac{k}{\delta} }$ to ensure that $f(t) \le N$ $\qed$. \\



\noindent \textbf{Proof of Lemma~\ref{lma:recursion_of_known_dt}.} 
The claim is trivial true when $N \in (0, 4k)$. If $4 k \le N \le 2 (\sqrt{2} + 2/3)^2 k \ln \frac{k}{\delta}$, then 
    \begin{align*}
         T(N) = T(N / 2) + O(k) 
            = T(4k) + O(k \log \frac{N}{4k}) 
            = O(k \log \frac{N}{k}) 
    \end{align*}
    Since $N \le 2 (\sqrt{2} + 2/3)^2 k \ln \frac{k}{\delta}$, we have $T(N) = O(k \log \log \frac{k}{\delta})$. 
    Finally, if $N > 2 (\sqrt{2} + 2/3)^2 k \ln \frac{k}{\delta}$, rewrite the following numbers as a power of two: 
    \begin{align*}
        (\sqrt{2} + 2/3)^2 k \ln \frac{k}{\delta}  = 2^{d_1}, 
        \quad 
        N = 2^{d_2} 
    \end{align*}
    for positive numbers $d_1 = \log \left( (\sqrt{2} + 2/3)^2 k \ln \frac{k}{\delta} \right)$ and $d_2 = \log N$. Define $S(d_2) = T(2^{d_2}) = T(N)= T\left((\sqrt{2} + 2/3)\sqrt{kN\ln \frac{k}{\delta}} \right) + O(k)$. Then 
    \begin{align*}
        S(d_2) 
	    &= S( (d_1 + d_2) / 2) + O(k) \\
            &= S( d_1 + (d_2 - d_1) / 2) + O(k) \\
            &= S( d_1 + (d_2 - d_1) / 4) + 2 \cdot O(k) \\ 
            &= ... \\
            &= S( d_1 + (d_2 - d_1) / 2^{\log (d_2 - d_1)} ) + \log (d_2 - d_1) \cdot O(k)
    \end{align*}
    By the definition of $S(\cdot)$, we have $S( d_1 + (d_2 - d_1) / 2^{\log (d_2 - d_1)} ) = S(d_1 + 1) = T(2^{d_1} \cdot 2) = T \left(2 (\sqrt{2} + 2/3)^2 k \ln \frac{k}{\delta} \right)$. Moreover, $\log (d_2 - d_1) = \log \log 2^{d_2 - d_1} = \log \log \frac{N}{(\sqrt{2} + 2/3)^2 k \ln \frac{k}{\delta}}$. Therefore, 
    \begin{equation*}
        T(N) = T \left(2 (\sqrt{2} + 2/3)^2 k \ln \frac{k}{\delta} \right) + O(k \log \log \frac{N}{k \ln \frac{k}{\delta}} )
    \end{equation*}
    The former term equals to $O(k \log \log \frac{k}{\delta})$. \myqed

    \vspace{2mm}
\noindent \textbf{Proof of Theorem~\ref{thm:correctness_of_dynamic_known_dt}}. Denote by $n_i$ the number of items received by player $i$ and by $I$ the set of players with $n_i > \mu_i t$ (i.e., the set of players that may send notifications to the coordinator). Our goal is to prove that players in $I$ send less than $k$ notifications\footnote{Without loss of generality, we assume that $\frac{N-t}{2k}$ is always an integer and thus, we can get rid of the floor operation. This is because otherwise, one can always use at most $O(k)$ straightforward communications to reduce $N-t$ to a multiple of $2k$. The communication bound will not be affected.}:
\begin{equation*}
    \sum_{i \in I}  \frac{n_i - \mu_i t}{(N - t) / 2k}  < k
\end{equation*}
First notice that by Lemma~\ref{lma:setting UB}, with probability at least $1 - \delta$, we have for all $i \in [k]$
\begin{equation*}
    |n_i - \mu_i t| \le \sqrt{2 t \mu_i (1 - \mu_i) \ln \frac{k}{\delta} } + \frac{2}{3} \ln \frac{k}{\delta}
\end{equation*}
by similar argument as Inequality~(\ref{ineq:concave_of_root_1}-\ref{ineq:concave_of_root_3}), we have 
\begin{align}
    \label{ineq:absolute_deviation_of_n_i}
    \sum_{i \in [k]} |n_i - \mu_i t| 
        &\le \sqrt{2 k t  \ln \frac{k}{\delta} } + \frac{2 k}{3} \ln \frac{k}{\delta} 
\end{align}
On the other hand, we have $\sum_{i = 1}^k n_i = t$, hence 
\begin{equation*}
    \sum_{i \in I} (n_i - \mu_i t) = \sum_{i \in [k]\setminus I} (\mu_i t - n_i)
\end{equation*}
It follows that 
\begin{align}
    \sum_{i \in [k]} |n_i - \mu_i t| 
    \label{eq:twice_excession}
    &= 2 \sum_{i \in I} (n_i - \mu_i t) 
\end{align}
Combining Inequality~(\ref{ineq:absolute_deviation_of_n_i}) and Equality~(\ref{eq:twice_excession}), we have 
\begin{align*}
    \sum_{i \in I}  \frac{n_i - \mu_i t}{(N - t) / 2k} 
        \le \frac{1}{2} \frac{ \sqrt{2 k t  \ln \frac{k}{\delta} } + \frac{2 k}{3} \ln \frac{k}{\delta} }{(N - t) / 2k} 
        < k
\end{align*}
The last inequality can be simplified to $t + \sqrt{2 k t  \ln \frac{k}{\delta} } + \frac{2 k}{3} \ln \frac{k}{\delta} < N$, which holds for $t = N - (\sqrt{2} + 2 / 3) \sqrt{k N \ln \frac{k}{\delta} }$, as proven in Lemma~\ref{lma:setting t}. 
\myqed

\noindent \textbf{Proof of Theorem~\ref{thm:time_cold_start}.} Denote $N_0$ the threshold to track when the algorithm begins and  let $w$ be the items tracked by $\cmy$ in the first round. As $\cmy$ tracks at least half the threshold in one round, it holds that $w \ge N_0 / 2$. After the first round, the threshold to track is $N \leftarrow N_0 - w$. Therefore, $w \ge N$ holds in all the subsequent rounds. 

For any subsequent round that runs our customized tracking algorithm, we are going to prove that: (i) it captures at least $t$ items, where $t$ is defined by Expression~(\ref{eqa:cold start t}); (ii) $\sum_{i \in [k]} UB_i \le N$; (iii) $t = N - O(\sqrt{kN \log \frac{k}{\delta}})$, in order to construct a similar recursion as the one that Lemma~\ref{lma:recursion_of_known_dt} solves. The theorem is proven by the same techniques used by Lemma~\ref{lma:recursion_of_known_dt}.

First, when the first $t$ items arrives, by Equation~(\ref{eqa:setting UB}), with probability $1 - \delta$, we have 
$$
n_i \le \mu_i t + \sqrt{2 t \mu_i (1-\mu_i) \ln \frac{k}{\delta} } + \frac{2}{3} \ln \frac{k}{\delta}
$$ 
for all $i \in [k]$. Observing that $(1 - \mu_i) \le 1$ and $\mu_i < \hat \mu_i$, we get
$$
n_i \le \hat \mu_i t + \sqrt{2 t \hat \mu_i \ln \frac{k}{\delta} } + \frac{2}{3} \ln \frac{k}{\delta}
$$ 
As a result, $UB_i = \hat \mu_i t + \sqrt{2 t \hat \mu_i \ln \frac{k}{\delta} } + \frac{2}{3} \ln \frac{k}{\delta}$ defined by Expression~(\ref{eqa:cold start upper}) is indeed an upper bound of $n_i$. 

Second, it remains to verify that the these $UB_i$ satisfy that \textit{correctness constraint} (Inequality~(\ref{ineq:correctness constraint})). Recall that $\sum_{i \in [k]} \hat \mu_i = \hat \Sigma$. Summing over $i \in [k]$, we have 
\begin{equation*}
    \sum_{i = 1}^k UB_i = t \hat \Sigma + \sum_{i = 1}^k \sqrt{2 t \hat \mu_i \ln \frac{k}{\delta} } + \sum_{i = 1}^k \frac{2}{3} \ln \frac{k}{\delta}
\end{equation*}
The third term sums up to $\frac{2}{3} k \ln \frac{k}{\delta}$. By concavity of the square root function, the second term is upper bounded by $\sqrt{2k t \hat \Sigma \ln \frac{k}{\delta} }$.
Now, by the definition of $ t = N / {\hat \Sigma} - (\sqrt{2} + \frac{2}{3} ) \sqrt{k (N / {\hat \Sigma}) \ln \frac{k}{\delta} }$ in Expression~(\ref{eqa:cold start t}), we have 
\begin{equation*}
    \sum_{i = 1}^k UB_i \le N - (\sqrt{2} + \frac{2}{3} ) \sqrt{k N {\hat \Sigma} \ln \frac{k}{\delta} } + \sqrt{2k N  \ln \frac{k}{\delta} } + \frac{2}{3} k \ln \frac{k}{\delta}
\end{equation*}
which concludes that $\sum_{i = 1}^k UB_i \le N$ as $N > k \ln \frac{k}{\delta}$ when this round runs our customized algorithm.

Finally, we need to show that $t = N - O(\sqrt{kN \log \frac{k}{\delta}})$. It suffices to show that $N /\hat \Sigma  = N - O(\sqrt{kN \log \frac{k}{\delta}})$. By substituting $\hat \mu_i$ with Expression~(\ref{eq:learned_upper}), and by a similar argument as in Inequality~(\ref{ineq:concave_of_root_1}-\ref{ineq:concave_of_root_3}), we get 
\begin{align*}
    \hat \Sigma
        &\le (1 + \sqrt{\frac{2 k  \ln \frac{3}{\delta} }{w} }  + \frac{3 k \ln \frac{3}{\delta} }{w} ) 
\end{align*}
Therefore, 
\begin{align*}
    N / \hat \Sigma \ge N / (1 + \sqrt{\frac{2 k  \ln \frac{3}{\delta} }{w} }  + \frac{3 k \ln \frac{3}{\delta} }{w} )
\end{align*}
Define $g(x) = \sqrt{\frac{2 k  \ln \frac{3}{\delta} }{x} }  + \frac{3 k \ln \frac{3}{\delta} }{x}$. By $w \ge N$, it holds $g(w) \le g(N)$. Hence $N(1 + g(w))(1 - g(N))\le N (1-g(N)^2) \le N$ and $N/\hat \Sigma \ge N / (1 + g(w)) \ge N(1 - g(N))$. Thus, the proof is completed. \myqed 



\section{Implementation Optimisations} \label{appendix:opt}

\noindent \textbf{Detecting non-stable distribution.} 
In general, the performance of a data-dependent algorithm may degenerate, when the empirical data does not follow the underlying distribution well.
This is also the case for our algorithms.
To remedy this issue, we propose a simple heuristic to detect whether the current empirical data still follows a distribution well.
If it does not, we switch to the $\cmy$ algorithm right away to minimize the impact of performance degeneration.
The heuristic works as follows.

Denote by $N$ the threshold to track at the start of the current round and $N'$ the one at the start of the next round. 
In other words, $N- N'$ items have been tracked in the current round. 
Intuitively, the ratio of $\frac{N - N'}{N}$ serves as an indicator of the effectiveness of the algorithm. 
If the ratio drops below a pre-specified threshold (say $0.75$), we switch the algorithm to the $\cmy$ algorithm. 
Such a mechanism guarantees that as soon as the empirical distribution is detected to be unstable, 
our algorithm will lose its effectiveness in at most one round, 
compared to the $\cmy$ algorithm. And therefore, at most $O(k)$ communication can be wasted and performance is still upper bounded by the communication bound of the $\cmy$ algorithm. 

\end{sloppy}

\end{document}